\documentclass[aps,pre,twocolumn,reprint,groupedaddress]{revtex4}
\usepackage{graphicx}
\usepackage{amsmath}
\usepackage{mathtools}
\usepackage{bbold}
\usepackage[usenames, dvipsnames]{color}

\usepackage{xcolor}

\begin{document}

\title{J1-J2 fractal studied by multi-recursion tensor-network method}

\author{Jozef \textsc{Genzor}$^{1}$}
\author{Andrej \textsc{Gendiar}$^{2},$\footnote{Corresponding author: andrej.gendiar@savba.sk}}
\author{Ying-Jer \textsc{Kao}$^{1}$}
\affiliation{$^1$Department of Physics, National Taiwan University, Taipei 10607, Taiwan}
\affiliation{$^2$Institute of Physics, Slovak Academy of Sciences, SK-845 11, Bratislava, Slovakia}

\date{\today}

\begin{abstract}

	We generalize a tensor-network algorithm to study thermodynamic properties of self-similar spin lattices constructed on a square-lattice frame with two types of couplings, $J_{1}^{}$ and $J_{2}^{}$, chosen to transform a regular square lattice ($J_{1}^{} = J_{2}^{}$) onto a fractal lattice if decreasing $J_{2}^{}$ to zero (the fractal fully reconstructs when $J_{2}^{} = 0$). We modified the Higher-Order Tensor Renormalization Group (HOTRG) algorithm for this purpose. Single-site measurements are performed by means of so-called impurity tensors. So far, only a single local tensor and uniform extension-contraction relations have been considered in HOTRG. We introduce ten independent local tensors, each being extended and contracted by fifteen different recursion relations. We applied the Ising model to the $J_{1}^{}-J_{2}^{}$ planar fractal whose Hausdorff dimension at $J_{2}^{} = 0$ is $d^{(H)} = \ln 12 / \ln 4 \approx 1.792$. The generalized tensor-network algorithm is applicable to a wide range of fractal patterns and is suitable for models without translational invariance.

\end{abstract}

\maketitle

\section{Introduction}

Understanding of phase transitions and critical phenomena plays an important role in condensed-matter physics. 
Much of the research on phase transitions has been devoted to the regular lattices.  
The two-dimensional classical Ising model on the square lattice is exactly solvable; on the other hand, there are no exact solutions for the spin models on fractals, so these must be studied numerically, requiring significantly higher effort than for regular lattices.

Many condensed matter systems can be characterized as fractal objects; for instance, percolation clusters, aggregates obtained from diffusion-limited growth processes, and adsorbent surfaces~\cite{Luscombe}.
Earlier studies on fractals from the viewpoint of the renormalization flow were carried out by Gefen et al.~\cite{Gefen1, Gefen2, Gefen3, Gefen4}. 
One of the main results yields the fact that the short-range classical spin models on finitely ramified lattices exhibit no phase transition at nonzero temperature~\cite{Gefen5, Luscombe}. 
The explanation can come from the relation of the boundary length of a finite-size fractal to its linear size, which strongly resembles one-dimensional systems.
However, the Ising model on Sierpi\'{n}ski carpet exhibits a phase transition~\cite{Vezzani}. 
There have been many attempts to study the Ising model on Sierpi\'{n}ski carpet numerically by Monte Carlo combined with the finite-size scaling method~\cite{Carmona, Monceau1, Monceau2, Pruessner, Bab}, including Monte Carlo renormalization group method~\cite{MCRG}.
Significant progress has been made recently in understanding the phase transition and critical phenomena on fractal lattices.

Adaptation of the Higher-Order Tensor Renormalization Group (HOTRG) for a fractal lattice with Hausdorff dimension $d^{(H)} = \ln 12 / \ln 4 \approx 1.792$ was introduced in Ref.~\onlinecite{2dising}. Therein, the numerical calculations were shown to be stable with respect to the bond dimension $D$. Based on the order parameter, the critical temperature was obtained, together with the critical magnetic exponent. The density matrix spectrum exhibits an exponential decay even at the critical point, which is possible to interpret as the system being less entangled because of the fractal geometry. The technical details of the methods used in~\cite{2dising} are presented in Ref.~\onlinecite{APS}. The hyper-scaling hypothesis is briefly discussed as well. Preliminary results on two separate infinite series of fractal lattices are presented and discussed.

In Ref.~\onlinecite{gasket}, HOTRG was applied to the study of the transverse-field Ising model on the Sierpi\'{n}ski fractal with the Hausdorff dimension $\log_2^{~} 3 \approx 1.585$. 
%
%
Ground/state energy and magnetization were calculated and analyzed.   

In Ref.~\onlinecite{carpet}, HOTRG was adapted to the classical Ising model on the Sierpi\'{n}ski carpet with the Hausdorff dimension $\log_3^{~} 8 \approx 1.8927$ using two types of local tensors. 
%
%
The position dependence of local thermodynamic functions was studied by employing impurity tensors, which were inserted at different locations on the fractal lattice. It was found that the critical exponent associated with the local spin polarization (spontaneous magnetization) varies by two orders of magnitude, depending on lattice location; however, the critical temperature $T_{\rm c}^{~}$ was found to be positionally independent.

Compared with their regular lattice counterparts, the geometrical details, such as lacunarity and connectivity, are the distinct key features of fractal lattices. 
If we embed a fractal lattice into a regular lattice and treat the coupling on the bonds not covered by the fractal differently, we can continuously interpolate between fractal and regular lattices. 
In this paper, we explore the phase transition phenomena on a particular family of lattices we can continuously interpolate between the planar fractal lattice~\cite{2dising} with the Hausdorff dimension $d^{(H)} = \ln 12 / \ln 4 \approx 1.792$ and the regular square lattice $d^{(H)} = 2$.
The property of self-similarity (i.~e., scale invariance) of the lattice is preserved throughout the transformation. 
However, only the regular square lattice is fully translationally invariant.

Here, we develop a numerically stable technique, which can also be applicable to a family of fractal lattices with partial translational non-invariance.
We call this family of lattices $J_1$-$J_2$ fractals, as our technique employs two types of couplings, $J_{1}^{}$ for bonds creating the fractal (thick bonds in black, as depicted in Fig.~\ref{fig:Fig_1}) and the remaining couplings $J_{2}^{}$ (thin bonds in red).  
%
%
A pure fractal lattice is obtained when specified bonds are cut by setting $J_{2}^{} = 0$, and a regular square lattice is recovered when $J_{1}^{} = J_{2}^{} = 1$. 

In order to analyze the phase transitions for the spin models on the fractal, we generalized the extension scheme in HOTRG~\cite{HOTRG}, as we have used in Refs.~\onlinecite{2dising, APS}.
Rather than considering only one type of the local tensor with the uniform extension relation, we introduce several types of local tensors (in this case ten), each one being extended by a different recursion relation. 
The recursion relations specify how to combine different tensors in order to extend the size of fractal iteratively, as required by HOTRG. 
The recursion relations reflect the symmetry of the self-similar lattices at every scale (i.~e., scale invariance) and are compatible with the framework of the renormalization group applied to tensor network states.
The computational cost scales with the bond dimension in the same way as in the two-dimensional HOTRG with a constant-factor overhead.

Local observables, such as magnetization and energy, can be implemented by means of impurity tensors. 
For concreteness, we focus on the Ising model on the fractal lattice shown in Fig.~\ref{fig:Fig_1}, which was recently studied by a different approach~\cite{2dising, APS} and is meant for comparison with the current approach, which has the potential to be applied to various types of fractals. 

We expect that the critical behavior may substantially change as the lattice transforms from the regular lattice to a fractal one. 
For example, if comparing the Ising model on the fractal and square lattices~\cite{APS, 2dising}, the numerically calculated critical temperatures $T_{\rm c}^{~}$ and associated magnetic critical exponents $\beta$ differed significantly, $T_{\rm c}^{~}\approx 1.31716$, $\beta \approx 0.0137$ and $T_{\rm c}^{~} = 2/\ln\left(1 + \sqrt{2}\right) = 2.26919...$, $\beta = 1/8 = 0.125$, respectively.

Moreover, we observed no divergence of the specific heat at the critical temperature in the fractal-lattice Ising model, as it has to be on the regular lattice.
Therefore, at some point during the transformation from the fractal to the regular lattice, 
the character of the phase transition must change from a weakly singular to the standard behavior, as known for the continuous (second-order) phase transition. 

\section{Model representation}  

\begin{figure}
\includegraphics[width=0.48\textwidth]{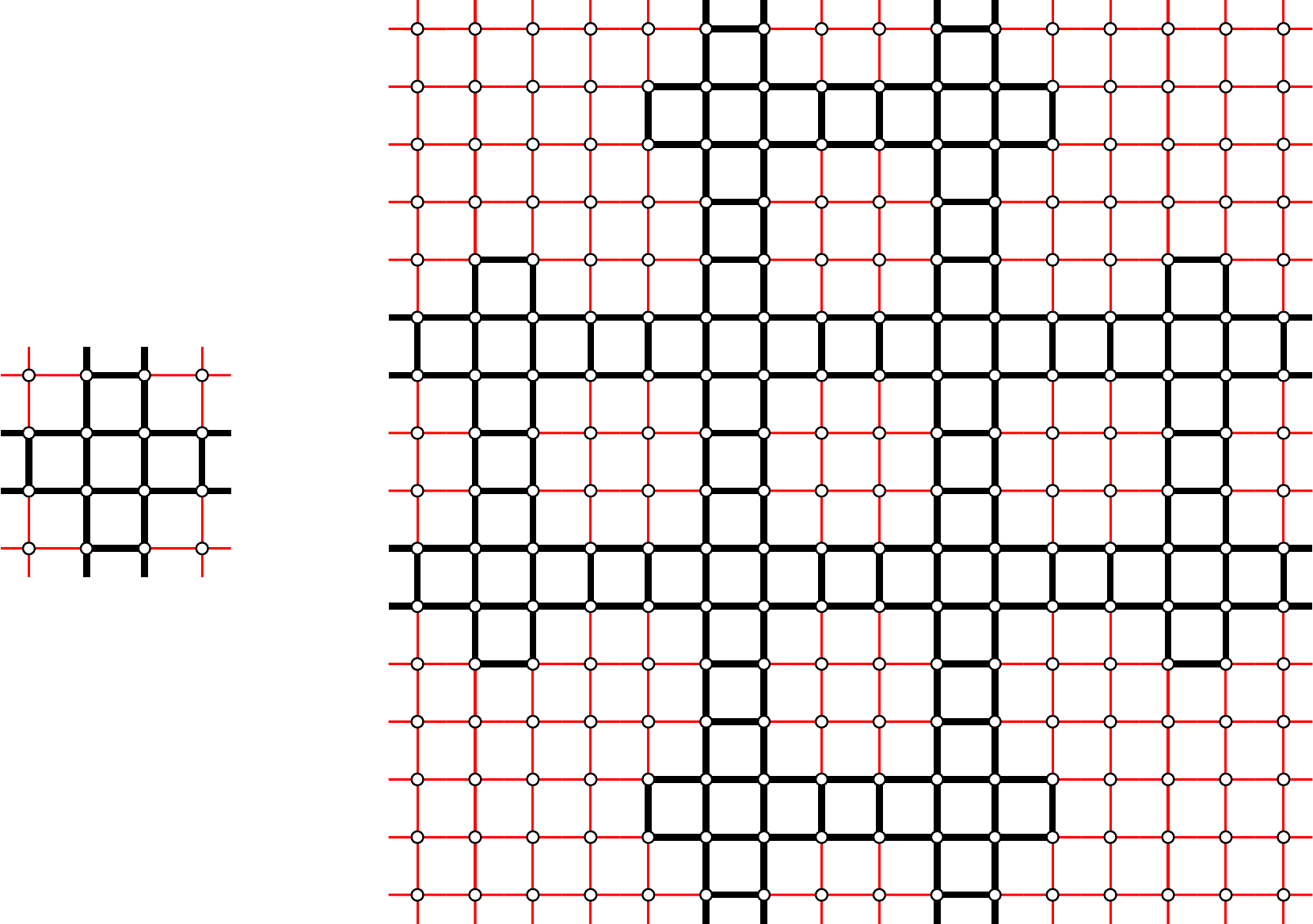}
\caption{
	One step of the growth process of the $J_1$-$J_2$ fractal lattice. Empty circles represent the two-state Ising spins. The thick (black) lines and the thin (red) lines represent the interactions with the spin-spin coupling equal to $J_1$ and $J_2$, respectively. Left: The basic $4\times4$ spin cluster is composed of $4^2$ vertices, where $12$ of them are connected by the $J_1$ coupling (in black), and the remaining four spin vertices are located in the corners only, being surrounded by the $J_2$ couplings (in red). Right: Extended cluster containing $16^2$ vertices with $12^2$ spin vertices connected via $J_1$ (the thick bonds in black) and the remaining 112 ($=4 * 16 + 12 * 4$) spin vortices with $J_2$ coupling only (thin bonds in red). 
}
\label{fig:Fig_1}
\end{figure}

First, we construct a square lattice with two types of bonds, the thick black ($J_1$) and the thin red ($J_2$), connecting only the nearest-neighbor lattice vertices, where the spins are located.
Hence, the fractal structure ($J_1$) and the remaining space ($J_2$) are comprised of black and red bonds, respectively. 
The iterative structure of HOTRG follows an extension series of how  to build up the (fractal) lattice. It starts from a unit cell made of $4 \times 4$ grid of spin vertices, where 12 vertices consist of three or four bonds $J_1$ (in black) and the remaining 4 vertices consist of four bonds $J_2$ placed in the corners, as in Fig.~\ref{fig:Fig_1} (left). 
After the initial 16-spin unit cell was copied 12-times, in the next iteration step, the identical pattern needs to be formed, as depicted for the $4 \times 4$ grid. It then becomes a $16 \times 16$ grid with the four corners, each made of the 16-spin vertices containing $J_2$ couplings only, in accord with Fig.~\ref{fig:Fig_1} (right). Notice that by disregarding the four corners, the expansion reduces to the process studied earlier~\cite{2dising}.
%

Consider the $J_1$-$J_2$ fractal Ising model Hamiltonian
\begin{equation}
{\cal H} = - J_1 \sum_{{\left<i j\right>}_1}^{~} \sigma_i \sigma_j - J_2 \sum_{{\left<i j\right>}_2}^{~} \sigma_i \sigma_j - h \sum_{i}^{~} \sigma_i \, ,
\end{equation}
where the Ising variable $\sigma$ takes the value $+1$ or $-1$, the non-negative ferromagnetic couplings $J_1$, $J_2$, and $h$ being the uniform external magnetic field. To distinguish the summation over $J_1$ and $J_2$ couplings, we separate the two sums and denote them ${\left< \right>}_1$ and ${\left< \right>}_2$, respectively. 
For brevity, we do not include the magnetic-field term $h$ in the following.
The two local Boltzmann weights (between two adjacent spins) are given by 
\begin{equation}
{\cal W}_{\text{B}}^{(\zeta)}\left(\sigma_i, \sigma_j\right) = \exp \left( \frac{J_{\zeta}}{k_{\rm B} T} \sigma_i \sigma_j \right) \, , 
\end{equation}
where the superscript index $\zeta = \{1, 2\}$ specifies the two types of the coupling $J_{\zeta}$. Here, $k_{\rm B}$ is the Boltzmann constant, and $T$ is temperature.
Without loss of generality, we will set $k_{\rm B}=1$ in what follows next.
The partition function is expressed as
\begin{equation}
{\cal Z} = \displaystyle\sum_{\{\sigma\}}
\prod_{{\left<i j\right>}_{\zeta}} {\cal W}_{\text{B}}^{(\zeta)} \left(\sigma_i, \sigma_j\right)
	\, ,
\end{equation}
where the sum is taken over all spin configurations $\{\sigma\}$. 
Furthermore, the bond weights ${\cal W}_{\text{B}}^{(\zeta)}$ can be re-expressed in terms of the matrix factorization
\begin{equation}
{\cal W}_{\text{B}}^{(\zeta)} \left( \sigma_i, \sigma_j \right) 
= \sum_{x=0}^{1} \, W_{\xi_{i} x_{~}}^{(\zeta)} \, W_{\xi_{j} x_{~}}^{(\zeta)} \, ,
\label{eKWW_fact}
\end{equation}
where the first matrix index $\xi_i=(1 - \sigma_i)/2$ takes the values $0$ or $1$ if $\sigma_i=\pm1$. Thus, the $2\times2$ matrix $W^{(\zeta)}$ has the elements
\begin{equation} \label{weight}
W^{(\zeta)} =  \left(\begin{array}{lr} 
\sqrt{\cosh J_{\zeta} /k_{\rm B} T} & \sqrt{\sinh J_{\zeta}/k_{\rm B} T}   \\
\sqrt{\cosh J_{\zeta} /k_{\rm B} T} & -\sqrt{\sinh J_{\zeta}/k_{\rm B} T} \end{array} \right) \, .
\end{equation}
Finally, we are ready to represent the partition function as a (non-homogeneous) tensor-network state at the $n^{\rm th}$ iteration step ($n=0,1,2,3,...$)
\begin{equation}
{\cal Z}_{n} = \text{Tr} \, \prod_{i_k} \, {\cal T}^{[k], n}_{x_i^{~} x_i^{\prime} y_i^{~} y_i^{\prime}}
\end{equation}
with the position-dependent local tensor ${\cal T}_{~}^{[k], n}$ specified at lattice site $i_k$. Each local tensor also carries information on types of couplings $J_1$ and $J_2$, which is sub-indexed by integer $[k]$ at the vertex position $i$. The four surrounded indices keep the fixed ordering so that $x_i^{~}$ (points to the left), $x_i^{\prime}$ (right), $y_i^{~}$ (up), and $y_i^{\prime}$ (down), i.e.,
\begin{equation}
{\cal T}^{[k], n}_{x_i^{~} x_i^{\prime} y_i^{~} y_i^{\prime}} = 
\sum_{\xi} 
W^{(\zeta)}_{\xi x_i^{~}} 
W^{(\zeta)}_{\xi x_i^{\prime}} 
W^{(\zeta)}_{\xi y_i^{~}} 
W^{(\zeta)}_{\xi y_i^{\prime}} \, ,
\end{equation}
where $\zeta=1$, $2$ specifies the coupling constant $J_{\zeta}$ depending on the orientation (left, right, up, down), which is determined by the index ordering in the position-dependent local tensor ${\cal T}$.

Notice that the local tensor can take up to $2^4$ different configurations (four indices/legs of the two states). On the other hand, the tensor-network state of the current fractal structure is constructed by ten types of the local tensors ${\cal T}_{~}^{[k], n=0}$, where $k=1, 2, 3, \dots , 10$, as in Tab.~\ref{table:Fig_2} (upper row). We initialize the local tensors at the zeroth iteration step ($n=0$) as
\begin{eqnarray}
\nonumber
{\cal T}_{x_{~}^{~} x_{~}^{\prime} y_{~}^{~} y_{~}^{\prime}}^{[1], {n=0}} &=& 
\sum_{\sigma} 
W^{(2)}_{\sigma x_{~}^{~}} 
W^{(2)}_{\sigma x_{~}^{\prime}} 
W^{(2)}_{\sigma y_{~}^{~}} 
W^{(2)}_{\sigma y_{~}^{\prime}} \, , \\
\nonumber
{\cal T}_{x_{~}^{~} x_{~}^{\prime} y_{~}^{~} y_{~}^{\prime}}^{[2], {n=0}} &=& 
\sum_{\sigma} 
W^{(2)}_{\sigma x_{~}^{~}} 
W^{(2)}_{\sigma x_{~}^{\prime}} 
W^{(1)}_{\sigma y_{~}^{~}} 
W^{(2)}_{\sigma y_{~}^{\prime}} \, , \\
\nonumber
{\cal T}_{x_{~}^{~} x_{~}^{\prime} y_{~}^{~} y_{~}^{\prime}}^{[3], {n=0}} &=& 
\sum_{\sigma} 
W^{(2)}_{\sigma x_{~}^{~}} 
W^{(1)}_{\sigma x_{~}^{\prime}} 
W^{(2)}_{\sigma y_{~}^{~}} 
W^{(2)}_{\sigma y_{~}^{\prime}} \, , \\
\nonumber
{\cal T}_{x_{~}^{~} x_{~}^{\prime} y_{~}^{~} y_{~}^{\prime}}^{[4], {n=0}} &=& 
\sum_{\sigma} 
W^{(2)}_{\sigma x_{~}^{~}} 
W^{(2)}_{\sigma x_{~}^{\prime}} 
W^{(2)}_{\sigma y_{~}^{~}} 
W^{(1)}_{\sigma y_{~}^{\prime}} \, , \\
\nonumber
{\cal T}_{x_{~}^{~} x_{~}^{\prime} y_{~}^{~} y_{~}^{\prime}}^{[5], {n=0}} &=& 
\sum_{\sigma} 
W^{(1)}_{\sigma x_{~}^{~}} 
W^{(2)}_{\sigma x_{~}^{\prime}} 
W^{(2)}_{\sigma y_{~}^{~}} 
W^{(2)}_{\sigma y_{~}^{\prime}} \, , \\
\label{T1}
{\cal T}_{x_{~}^{~} x_{~}^{\prime} y_{~}^{~} y_{~}^{\prime}}^{[6], {n=0}} &=& 
\sum_{\sigma} 
W^{(2)}_{\sigma x_{~}^{~}} 
W^{(1)}_{\sigma x_{~}^{\prime}} 
W^{(1)}_{\sigma y_{~}^{~}} 
W^{(2)}_{\sigma y_{~}^{\prime}} \, , \\
\nonumber
{\cal T}_{x_{~}^{~} x_{~}^{\prime} y_{~}^{~} y_{~}^{\prime}}^{[7], {n=0}} &=& 
\sum_{\sigma} 
W^{(2)}_{\sigma x_{~}^{~}} 
W^{(1)}_{\sigma x_{~}^{\prime}} 
W^{(2)}_{\sigma y_{~}^{~}} 
W^{(1)}_{\sigma y_{~}^{\prime}} \, , \\
\nonumber
{\cal T}_{x_{~}^{~} x_{~}^{\prime} y_{~}^{~} y_{~}^{\prime}}^{[8], {n=0}} &=& 
\sum_{\sigma} 
W^{(1)}_{\sigma x_{~}^{~}} 
W^{(2)}_{\sigma x_{~}^{\prime}} 
W^{(2)}_{\sigma y_{~}^{~}} 
W^{(1)}_{\sigma y_{~}^{\prime}} \, , \\
\nonumber
{\cal T}_{x_{~}^{~} x_{~}^{\prime} y_{~}^{~} y_{~}^{\prime}}^{[9], {n=0}} &=& 
\sum_{\sigma} 
W^{(1)}_{\sigma x_{~}^{~}} 
W^{(2)}_{\sigma x_{~}^{\prime}} 
W^{(1)}_{\sigma y_{~}^{~}} 
W^{(2)}_{\sigma y_{~}^{\prime}} \, , \\
\nonumber
{\cal T}_{x_{~}^{~} x_{~}^{\prime} y_{~}^{~} y_{~}^{\prime}}^{[10], {n=0}} &=& 
\sum_{\sigma} 
W^{(1)}_{\sigma x_{~}^{~}} 
W^{(1)}_{\sigma x_{~}^{\prime}} 
W^{(1)}_{\sigma y_{~}^{~}} 
W^{(1)}_{\sigma y_{~}^{\prime}} \, .
\end{eqnarray}

\subsection{Coarse-graining procedure}

\begin{table*}[tb]
\begin{center}
\begin{tabular}{llllllllllll}
{\vspace{-0.0cm} $k$} & $1$ & $2$ & $3$ & $4$ & $5$ & $6$ & $7$ & $8$ & $9$ & $10$ \\ [0.2cm]
{\vspace{-0.0cm} ${\cal T}^{[k], n}$} & & & & & & & & & &  \\ [-0.4cm]
&
\includegraphics[width=0.59in]{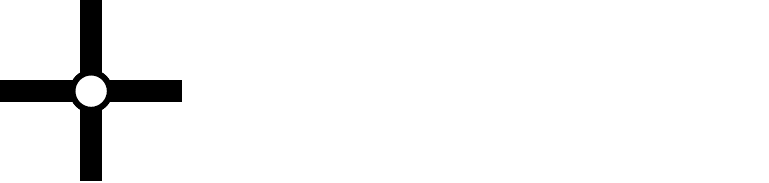} &
\includegraphics[width=0.59in]{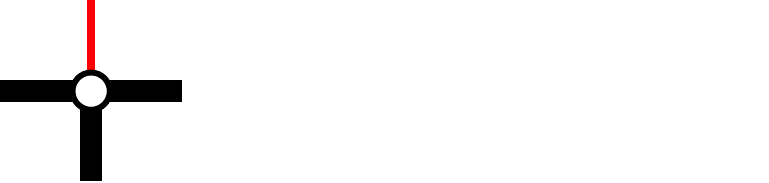} &
\includegraphics[width=0.59in]{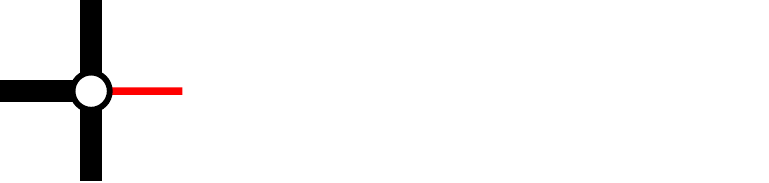} &
\includegraphics[width=0.59in]{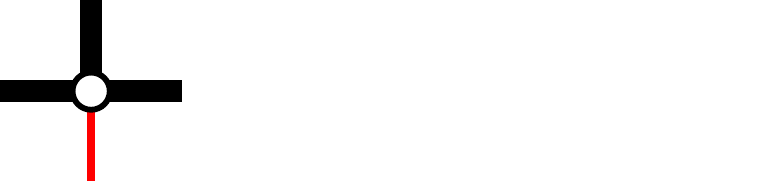} &
\includegraphics[width=0.59in]{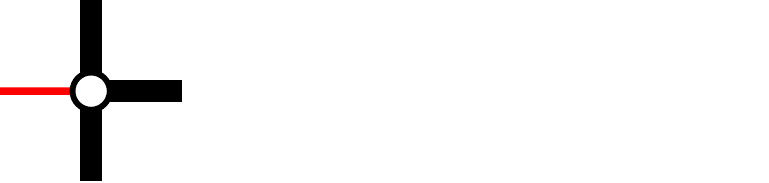} &
\includegraphics[width=0.59in]{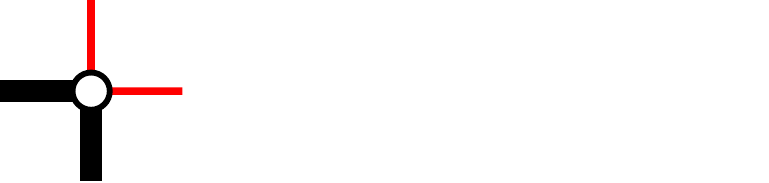} &
\includegraphics[width=0.59in]{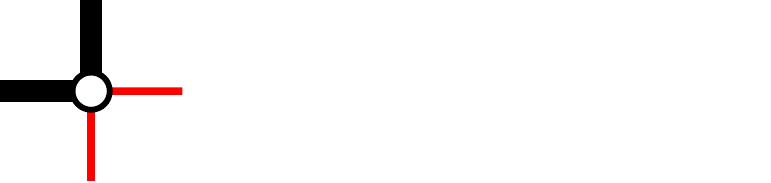} &
\includegraphics[width=0.59in]{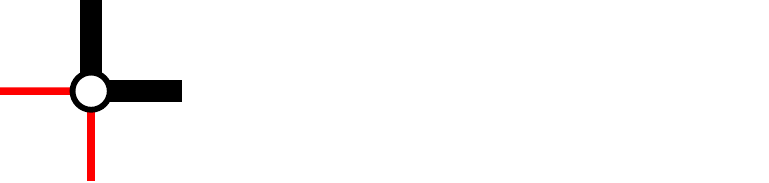} &
\includegraphics[width=0.59in]{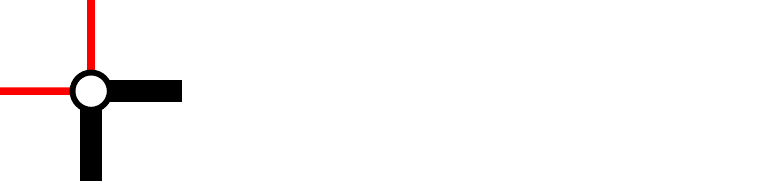} &
\includegraphics[width=0.59in]{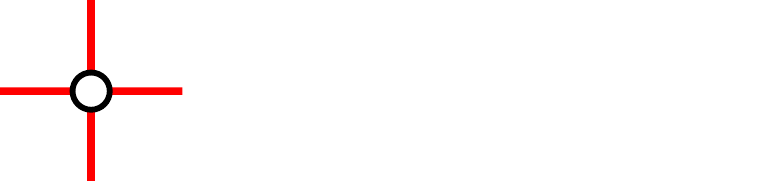} \\ [0.3cm]
{\vspace{-0.0cm} ${\cal T}^{[k], n+1}$} & & & & & & & & & &  \\ [-0.5cm]
&
\includegraphics[width=0.59in]{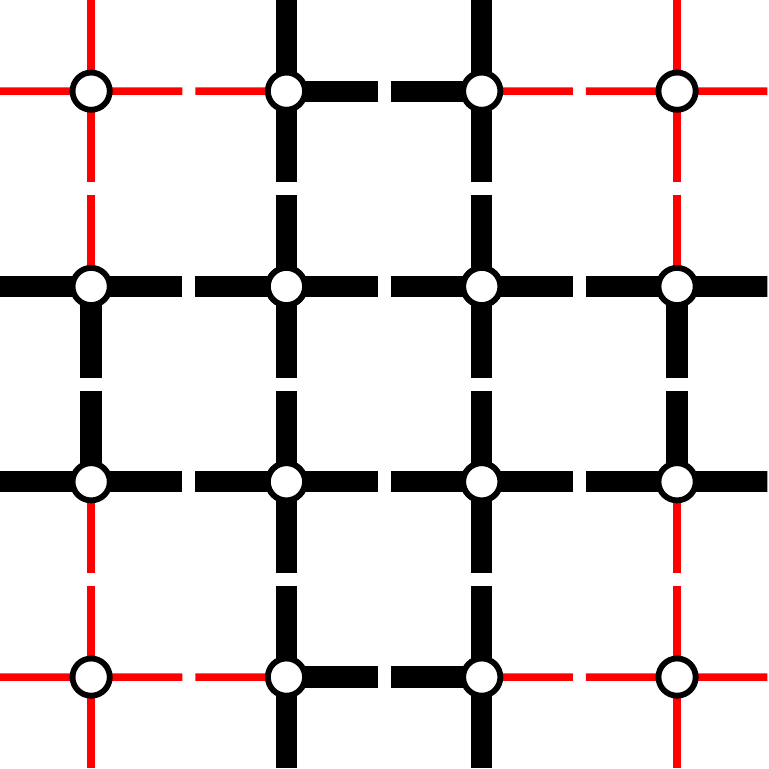} &
\includegraphics[width=0.59in]{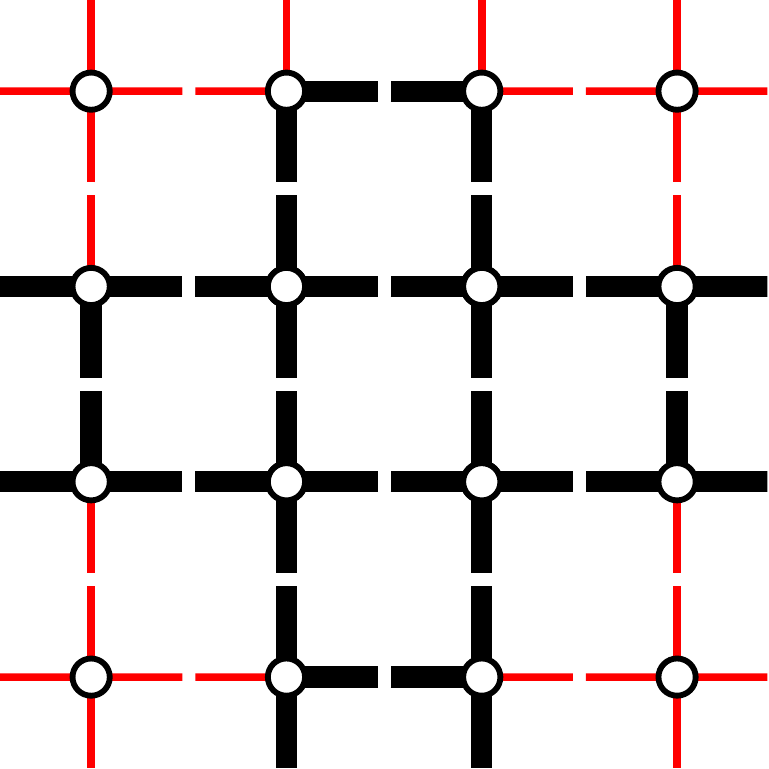} &
\includegraphics[width=0.59in]{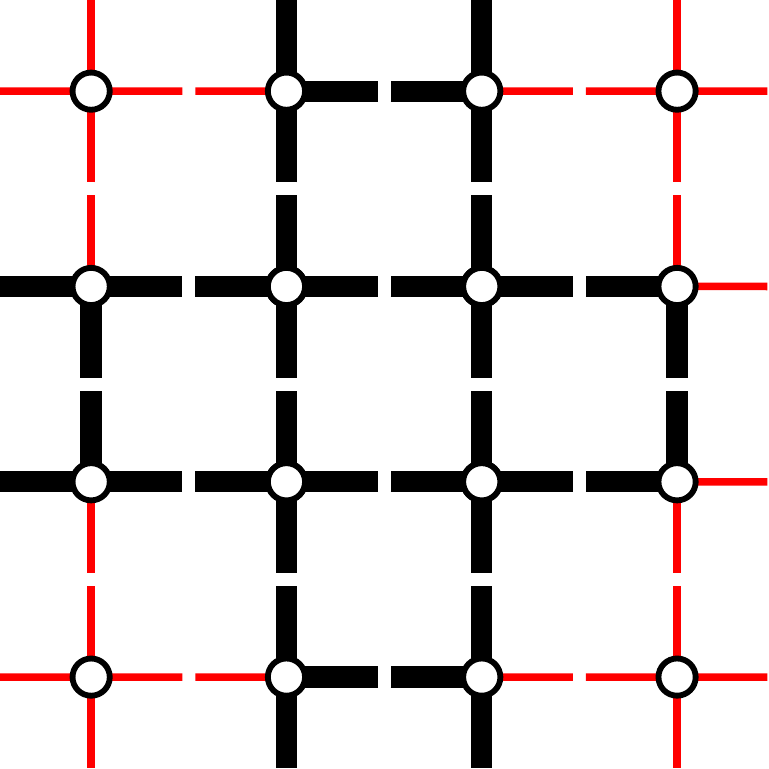} &
\includegraphics[width=0.59in]{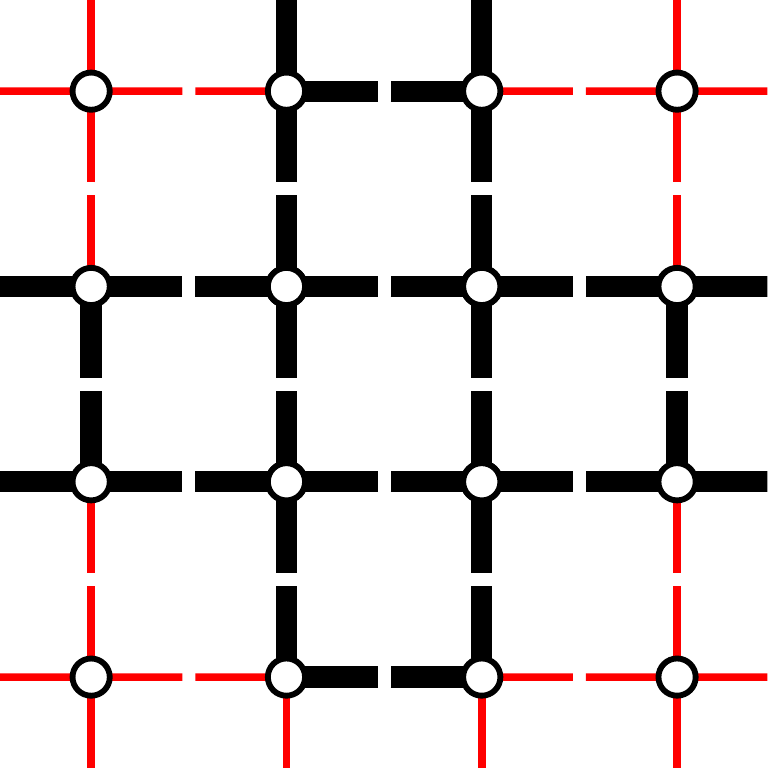} &
\includegraphics[width=0.59in]{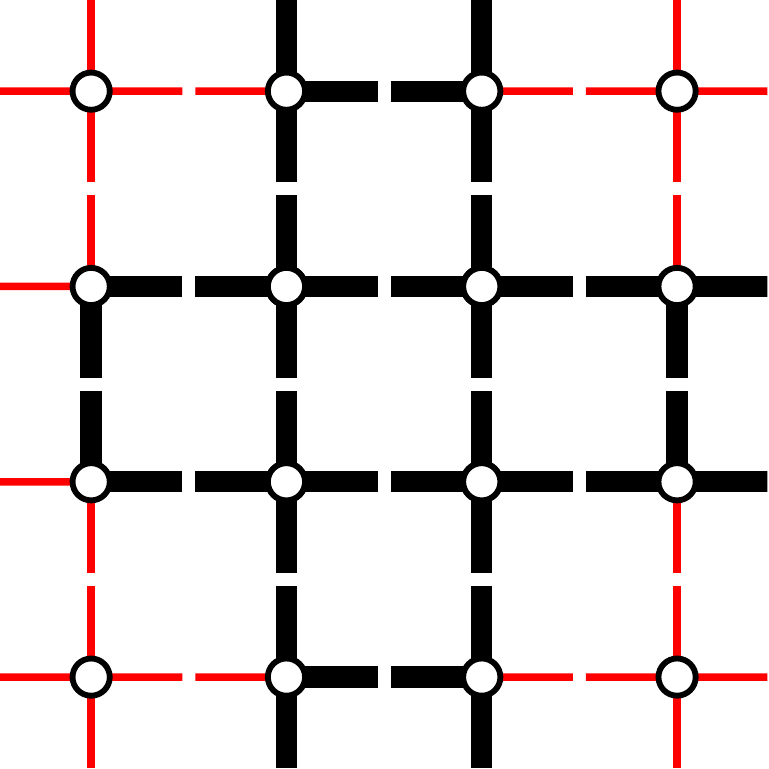} &
\includegraphics[width=0.59in]{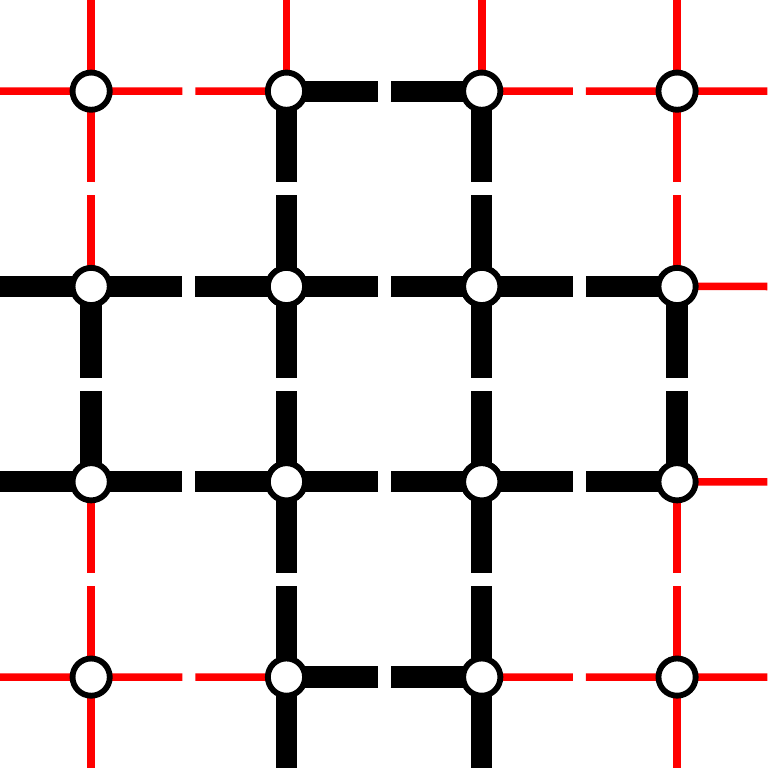} &
\includegraphics[width=0.59in]{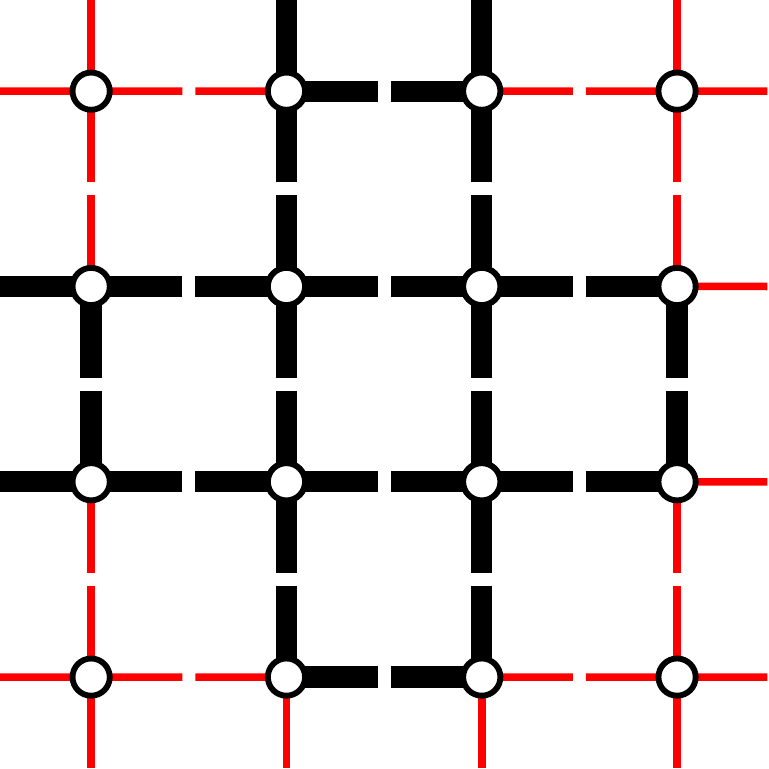} &
\includegraphics[width=0.59in]{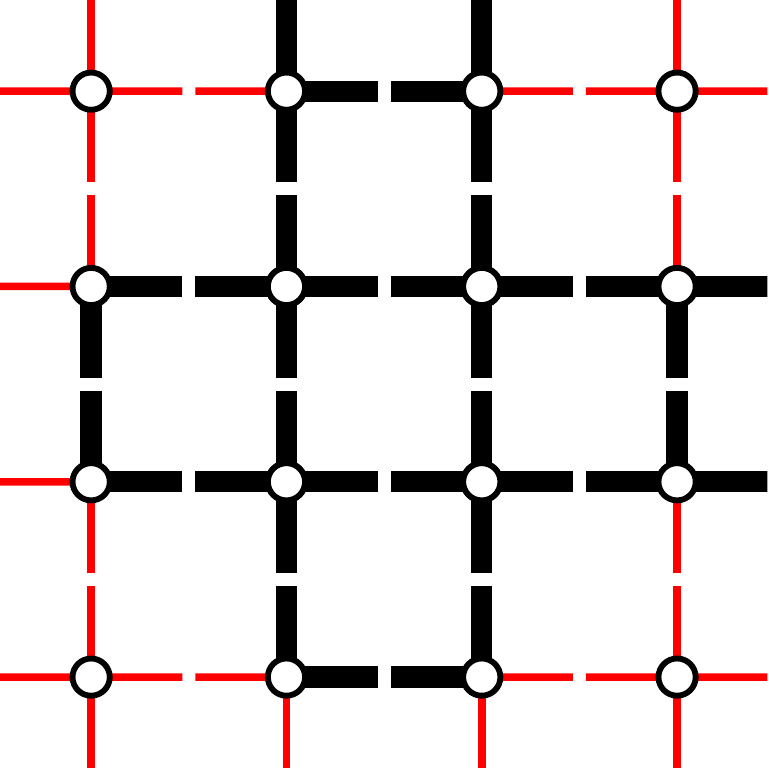} &
\includegraphics[width=0.59in]{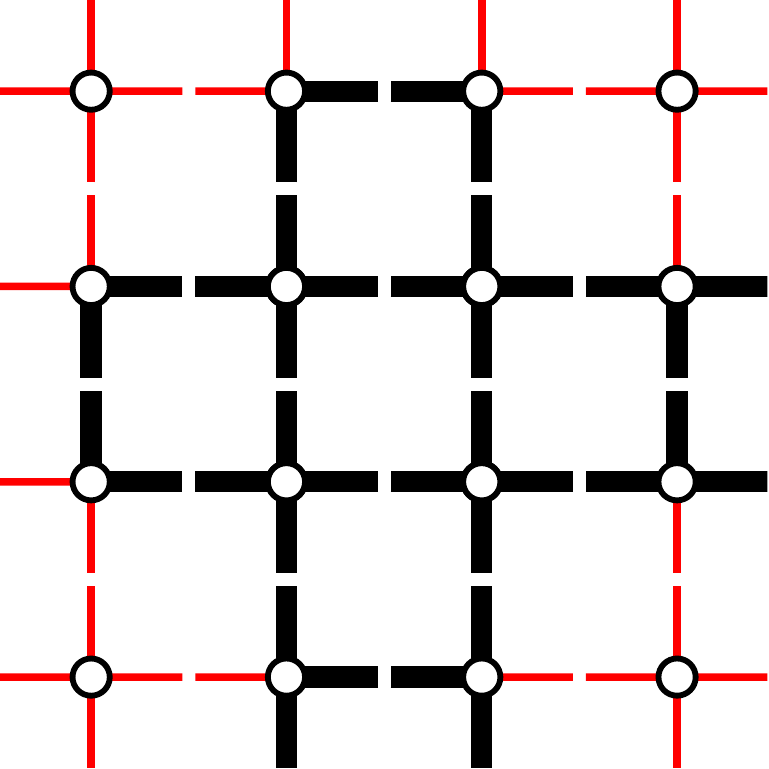} &
\includegraphics[width=0.59in]{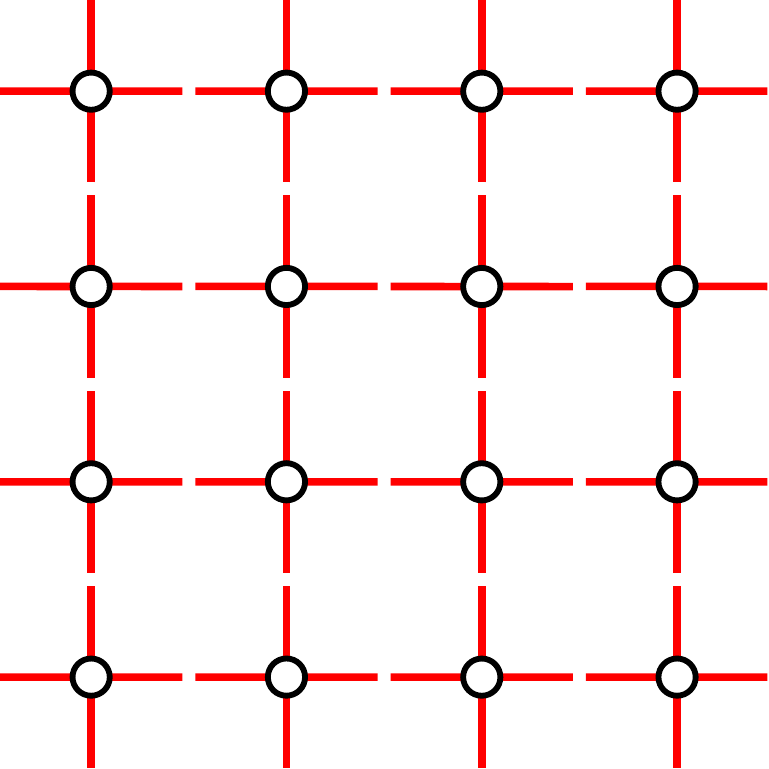}
\end{tabular}
\end{center}
 \caption{(Color online) 
The graphical representation of the ten local tensors ${\cal T}^{[k], n}$ showing the way of arrangement for the next iteration step.
	Upper row: Array of the local tensors at the $n^{\rm th}$ iteration step ${\cal T}^{[k], n}$.
	Lower row: The respective extension patterns specifying how the local tensors ${\cal T}^{[k], n}$ are combined to form the extended tensors ${\cal T}^{[k], n+1}$ for the next iteration step.
}
\label{table:Fig_2}
\end{table*}
%
%
%

To consistently define the iterative extension procedure, we need to extend each of the ten local tensors according to the specific extension relation for the next iteration step $n+1$, as graphically summarized in Tab.~\ref{table:Fig_2}.
After the extension procedure of the ten extended tensors ${\cal T}_{~}^{[k], n+1}$ is finalized, we will apply (renormalization) transformations to reduce degrees of freedom of the expanded tensors, specified later.
Multiple types of the local tensors enter the extension relation for each tensor. Therefore, all the extensions for the next iteration step $n+1$ need to be performed by means of those from the previous iteration step $n$, which have to be kept simultaneously, until the entire update of the tensors is completed. 
(The tensor ${\cal T}_{~}^{[10]}$ is a special case and is extended with the copies of itself.) 

As a typical example, consider the extension of the tensor ${\cal T}_{~}^{[1]}$ in detail. According to Tab.~\ref{table:Fig_2}, for obtaining the new tensor ${\cal T}_{~}^{[1], n+1}$, we contract 16 tensors at step $n$ in total (four tensors of type $k=1$, two of type $k=2$, two of type $k=3$, two of type $k=4$, two of type $k=5$, and, finally, four of type $k=10$) arranged onto a $4\times4$ grid to satisfy the extended pattern $k=1$ for the next step $n+1$
\begin{gather}
\label{Eq_Crazy}
{\cal T}_{
\left(x_1^{~}x_2^{~}x_3^{~}x_4^{~}\right) 
\left(x_{1}^{\prime}x_{2}^{\prime}x_{3}^{\prime}x_{4}^{\prime}\right) 
\left(y_1^{~}y_2^{~}y_3^{~}y_4^{~}\right) 
\left(y_{1}^{\prime}y_{2}^{\prime}y_{3}^{\prime}y_{4}^{\prime}\right)}^{[1], {n+1}}\\
=\quad \raisebox{-3.2em}{\includegraphics[height=7em]{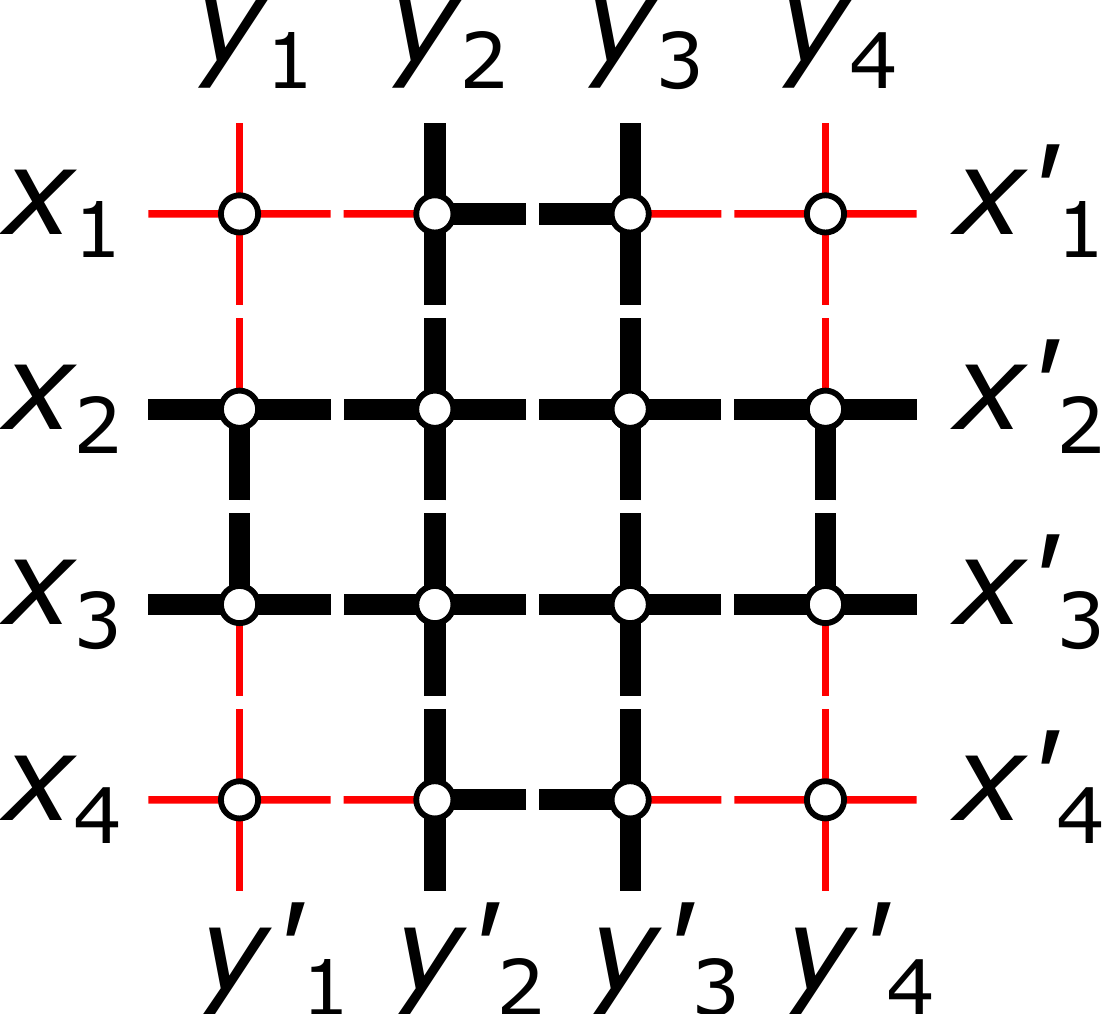}} 
 \, .  \nonumber
\end{gather}
Analogous extension relations hold for the remaining tensors, as listed in Tab.~\ref{table:Fig_2} and more details can be found in Appendix~\ref{app_A}.

\section{Renormalization transformation}

\begin{figure}
\includegraphics[width=0.48 \textwidth]{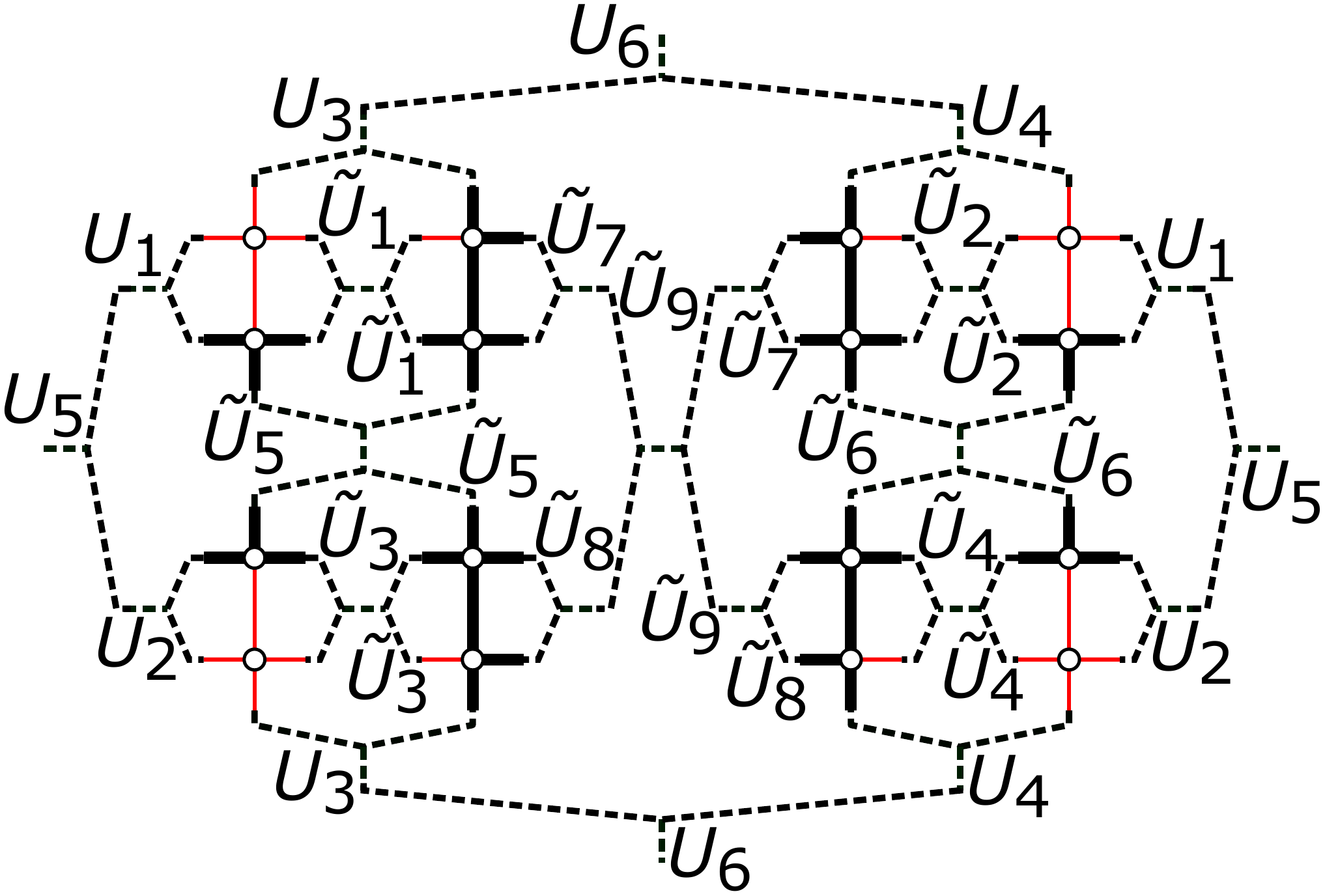}
\caption{ (Color online)
Graphical visualization of the extension and renormalization procedures to build up the tensor ${\cal T}^{[1],n+1}$ out of the appropriate tensors ${\cal T}^{[k],n}$. 
In total, 15 different projectors are introduced to perform four steps of the coarse-graining in HOTRG. 
	We stress the way of distinguishing the {\it external} projectors ($U_{l}^{~}$, $l=1,2,\dots, 6$) which perform the projections onto the {\it external} legs from the {\it internal} projectors ($\tilde{U}_{l}^{~}$, $l=1,2,\dots, 9$) which perform the {\it internal} projections within the $4\times4$ grid. 
}
\label{renorm}
\end{figure}

One step of the coarse-gaining procedure defined by Eq.~\eqref{Eq_Crazy} increases the bond dimension as the fourth power. It is, thus, numerically inefficient to exactly contract all of the 16 tensors comprising one unit cell in just a single step. A simple way of introducing an efficient approximation is to perform four steps of the HOTRG-style coarse-graining by contracting neighboring pairs of tensors at each step while specifying all the adjacent projectors, being properly matched, as depicted in Fig.~\ref{renorm}. Notice Fig.~\ref{renorm}, where we have introduced 15 different projectors, where six of the projectors ($U_{l}^{~}$, $l=1,2,\dots, 6$) perform projections onto the {\it external} (renormalized) tensor indices/legs, whereas nine of the projectors ($\tilde{U}_{l}^{~}$, $l=1,2,\dots, 9$) perform {\it internal} projections inside the $4\times4$ tensor grid. 

Now, let us demonstrate how the internal renormalization transformations $\tilde{U}_1$ and $\tilde{U}_2$ are calculated. By contracting the tensors ${\cal T}^{[5], n}$ and ${\cal T}^{[1], n}$ along the $y$ axis, we define
\begin{equation}
M^{[5, 1], n}_{x x^{\prime} y y^{\prime}} = \sum_i {\cal T}^{[5], n}_{x_1^{~} x_1^{\prime} y_{~}^{~} i_{~}^{~}}  {\cal T}^{[1], n}_{x_2^{~} x^{\prime}_2 i_{~}^{~} y_{~}^{\prime}} \, ,
\end{equation}
where $x = x_1 \otimes x_2$ and $x' =x'_1 \otimes x'_2$.
To truncate the tensor $M^{[5, 1], n}$ in accord with the higher-order singular value decomposition (HOSVD)~\cite{hosvd}, the following matrix unfolding has to be prepared
\begin{equation} \label{unfolding_1}
{M^{\prime}}^{[5, 1], n}_{x \left(x^{\prime} y y^{\prime}\right)} = M^{[5, 1], n}_{x x^{\prime} y y^{\prime}} \, , 
\end{equation}
We then perform the singular-value decomposition (SVD)
\begin{equation} \label{svd1}
{M^{\prime}}^{[5, 1], n}_{~} = \tilde{U}_{1}^{~} \tilde{\omega}_{1}^{~} \tilde{V}_{1}^{\dag} \, ,
\end{equation}
where $\tilde{U}_{1}^{~}$ and $\tilde{V}_{1}^{\dag}$ are unitary matrices of the respective dimensions, and $\tilde{\omega}_{1}^{~}$ is a diagonal matrix with the non-negative singular values on its diagonal ordered in the decreasing order by convention.

To obtain the {\it internal} renormalization transformation $\tilde{U}_2$, we contract the tensors ${\cal T}^{[3], n}$ and ${\cal T}^{[1], n}$ along the $y$ axis
\begin{equation}
M^{[3, 1], n}_{x x^{\prime} y y^{\prime}} = \sum_i {\cal T}^{[3], n}_{x_1^{~} x_1^{\prime} y_{~}^{~} i_{~}^{~}}  {\cal T}^{[1], n}_{x_2^{~} x^{\prime}_2 i_{~}^{~} y_{~}^{\prime}} \, ,
\end{equation}
where $x = x_1 \otimes x_2$ and $x' =x'_1 \otimes x'_2$.
To truncate the tensor $M^{[3, 1], n}$ by HOSVD, the following matrix unfolding is prepared
\begin{equation}
{M^{\prime}}^{[3, 1], n}_{x^{\prime} \left(y y^{\prime} x\right)} = M^{[3, 1], n}_{x x^{\prime} y y^{\prime}} \, . 
\end{equation}
Notice that this unfolding is different from Eq.~\eqref{unfolding_1}, as we optimize the right side of the contracted tensor $M^{[3, 1], n}$, as opposed to the left side.
From SVD we get
\begin{equation} \label{svd2}
{M^{\prime}}^{[3, 1], n}_{~} = \tilde{U}_{2}^{~} \tilde{\omega}_{2}^{~} \tilde{V}_{2}^{\dag} \, ,
\end{equation}
where $\tilde{U}_{2}^{~}$ and $\tilde{V}_{2}^{\dag}$ are unitary matrices, and $\tilde{\omega}_{2}^{~}$ is a diagonal matrix with singular values §(ordered decreasingly). 
The remaining projectors are calculated similarly.

The singular values obtained by SVD in Eq.~\eqref{svd1} or Eq.~\eqref{svd2} can be used to calculate the entanglement entropy. 
Alternatively, we define the entanglement entropy using the singular values obtained from SVD applied directly to the tensor ${\cal T}^{[1], n}$ unfolded into a matrix as ${{\cal T}^{\prime}}^{[1], n}_{x \left(x^{\prime} y y^{\prime}\right)} = {\cal T}^{[1], n}_{x x^{\prime} y y^{\prime}}$. Having performed the SVD ${T^{\prime}}^{[1], n}_{~} = U_{~}^{~} \omega_{~}^{~} V_{~}^{\dag}$,
%
%
we calculate the entanglement entropy $s$ as follows
\begin{equation} \label{entanglement}
	s^{~}_{~} = - {\rm Tr}\ \rho \ln \rho = - \displaystyle\sum_{\xi=1}^{D} \dfrac{\left(\omega_{\xi}\right)^2}{\Omega} \ln \dfrac{\left(\omega_{\xi}\right)^2}{\Omega} \, ,
\end{equation}
where $\rho$ is a reduced density matrix and $\Omega = \sum_{\xi=1}^{D} (\omega_{\xi}^{~})^2$ is a normalization factor such that ${\rm Tr} \rho = 1$.

The projectors $U$ and $\tilde{U}$ are obtained with the extension pattern ($k=1$) corresponding to ${\cal T}^{[1]}$. However, we can consider applying these projectors uniformly to the remaining extension patterns with $k>1$. With such a simple setup, the projectors' consistency at the boundaries between all the unit cells is clearly satisfied. If necessary, the {\it internal} projectors $\tilde{U}$ can be obtained from each extension pattern separately, which can increase numerical accuracy a bit, for the price of higher computational costs (by a constant factor at most). The {\it external} projectors $U$, however, have to be uniform, and we need to decide which of the projectors to apply at the boundaries of the unit cells.

We have encountered some numerical instabilities when projecting the tensor patterns with $k \geq 2$ using the projectors obtained from ${\cal T}^{[1]}$.
To improve the approximation scheme described above, we can introduce multiple sets of the {\it external} projectors. As a simple yet practical example, let us consider two sets of {\it external} projectors $U^{1}_{l}$, $U^{2}_{l}$ ($l=1,2,\dots,6$), each containing six HOTRG isometries. The projectors $U^{1}_{l}$ are obtained as before (i.~e. on ${\cal T}^{[1]}$); however, $U^{2}_{l}$ are obtained on the homogeneous pattern defining ${\cal T}^{[10]}$. We can then use $U^{1}$ when truncating a thick leg (black) and $U^{2}$ when truncating a thin leg (red). 
For instance, the renormalization relations of the tensors ${\cal T}^{[1]}$, ${\cal T}^{[2]}$, ${\cal T}^{[6]}$, and ${\cal T}^{[10]}$ for the {\it external} legs are straightforward
\begin{gather} 
\label{R1}
{\cal T}^{[1], n+1}_{x_{~}^{~} x_{~}^{\prime} y_{~}^{~} y_{~}^{\prime}} = \quad \raisebox{-4.3em}{\includegraphics[height=9em]{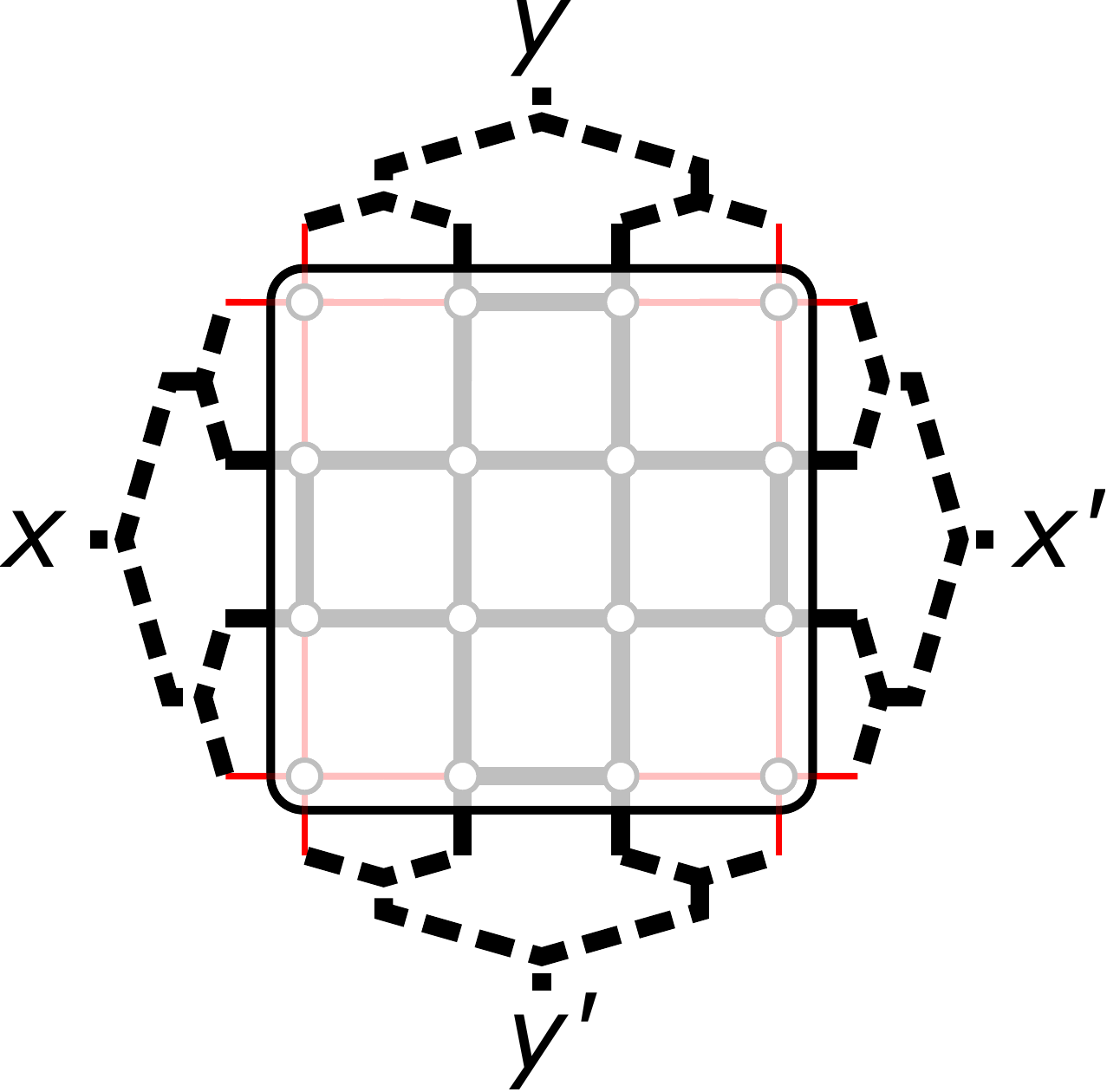}} 
 \, ,
\end{gather}
\begin{gather} 
\label{R2}
{\cal T}^{[2], n+1}_{x_{~}^{~} x_{~}^{\prime} y_{~}^{~} y_{~}^{\prime}} = \quad \raisebox{-4.3em}{\includegraphics[height=9em]{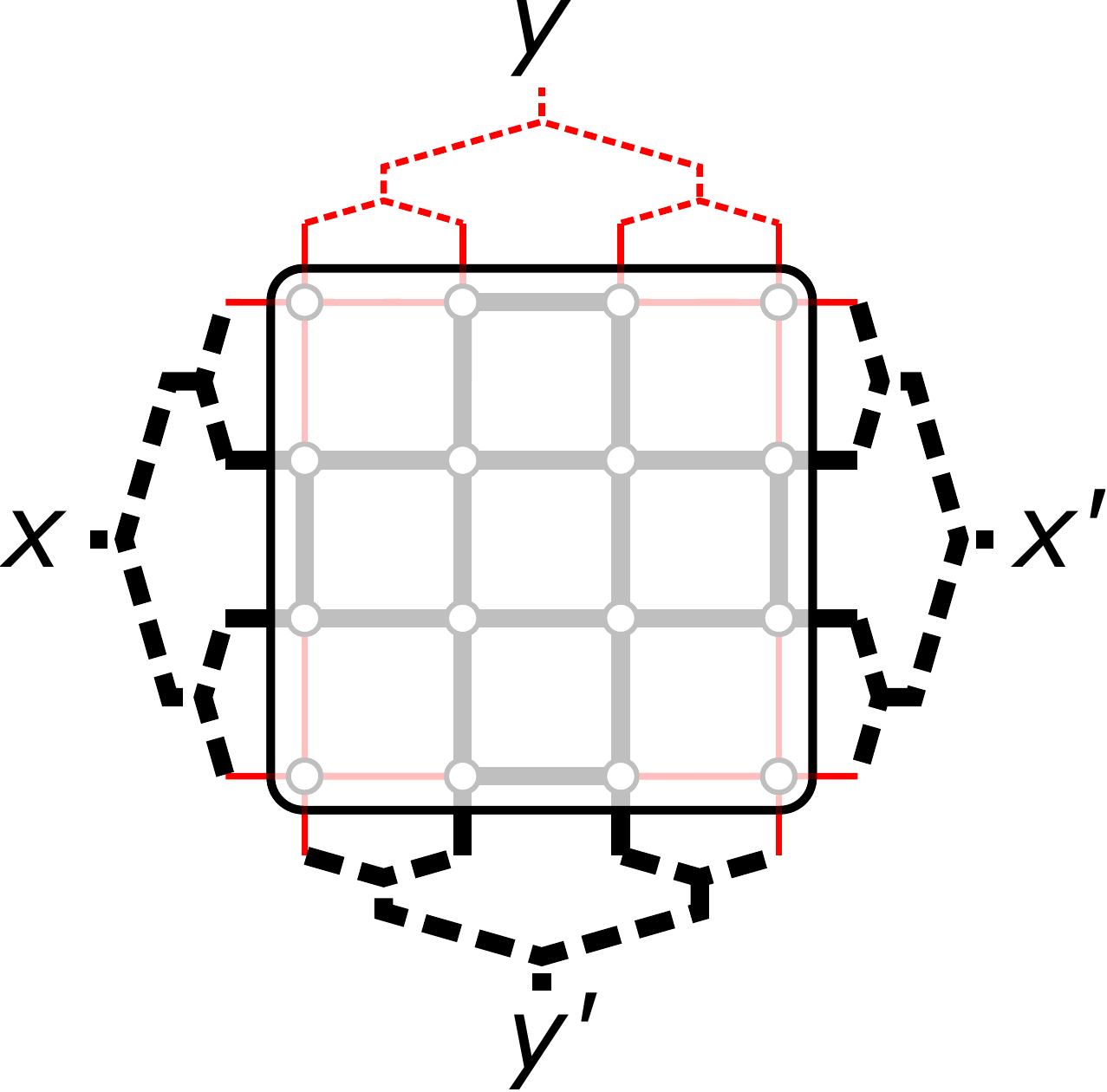}} 
 \, ,
\end{gather}
\begin{gather} 
\label{R3}
{\cal T}^{[6], n+1}_{x_{~}^{~} x_{~}^{\prime} y_{~}^{~} y_{~}^{\prime}} = \quad \raisebox{-4.3em}{\includegraphics[height=9em]{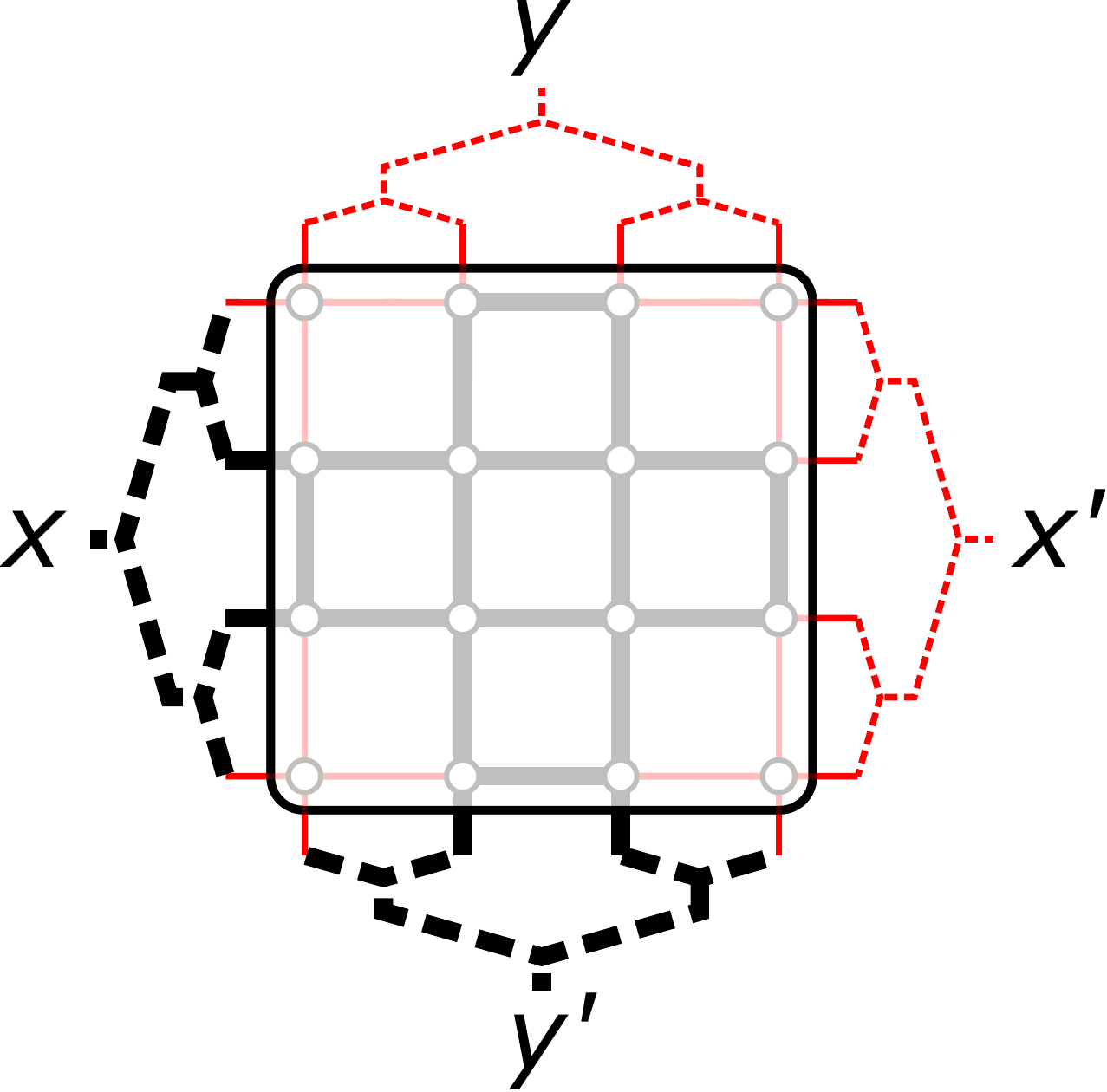}} 
 \, , 
\end{gather}
\begin{gather} 
\label{R4}
{\cal T}^{[10], n+1}_{x_{~}^{~} x_{~}^{\prime} y_{~}^{~} y_{~}^{\prime}} = \quad \raisebox{-4.3em}{\includegraphics[height=9em]{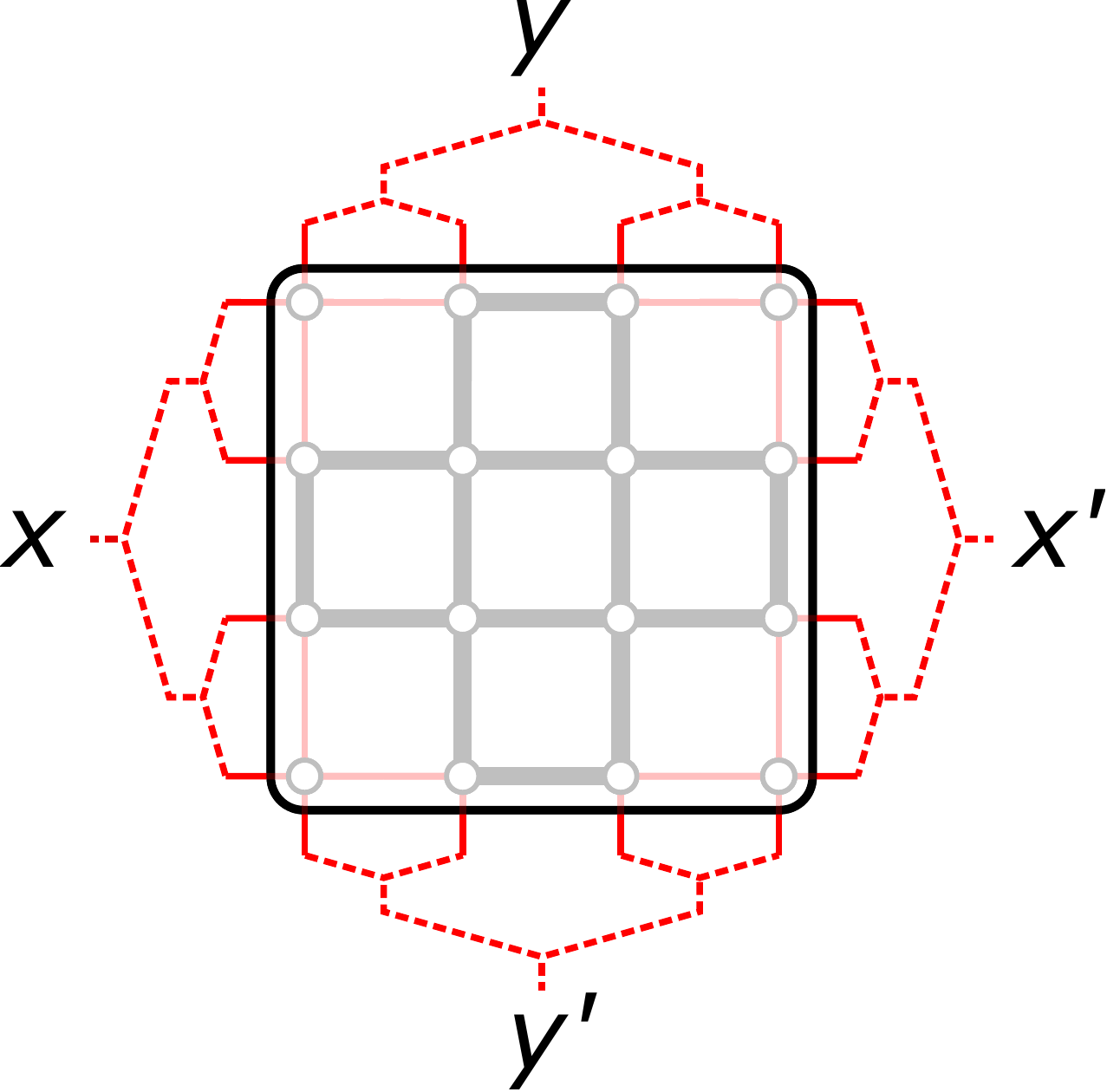}} 
 \, , 
\end{gather}
where the projectors $U^{1}$ and $U^{2}$ are depicted by dashed thick (black) and thin (red) lines, respectively.
The explicit form of Eqs.~\eqref{R1}--\eqref{R4} is presented in Appendix~\ref{app_B}.
For completeness, we list all the projection patterns in Appendix~\ref{app_B} as well.

\subsection{Impurity tensors}

\noindent
{\it Magnetization}: We can define the impurity tensor $\tilde{\cal T}^{n=0}$ by inserting a spin variable $\sigma = 1 - 2 \xi$ into the local tensor with the pattern ${\cal T}^{[1],n=0}$ as follows (cf.~Eq.~(\ref{T1}))
\begin{equation} \label{mag_imp_init}
\tilde{\cal T}_{x_i^{~} x_i^{\prime} y_i^{~} y_i^{\prime}}^{n=0} = 
\sum_{\xi} \left( 1 - 2 \xi \right)
W^{(2)}_{\xi x_{~}^{~}} 
W^{(2)}_{\xi x_{~}^{\prime}} 
W^{(2)}_{\xi y_{~}^{~}} 
W^{(2)}_{\xi y_{~}^{\prime}} \, .
\end{equation}
The extension of the impurity tensor, $\tilde{\cal T}^{n}\to\tilde{\cal T}^{n+1}$, is performed by taking an average over four central vertices in the extension pattern ${\cal T}^{[1],n+1}$  (cf.~Eq.~(\ref{Eq_Crazy}))
\begin{eqnarray} \label{extension_imp1}
\tilde{\cal T}_{
\left(x_1^{~}x_2^{~}x_3^{~}x_4^{~}\right) 
\left(x_{1}^{\prime}x_{2}^{\prime}x_{3}^{\prime}x_{4}^{\prime}\right) 
\left(y_1^{~}y_2^{~}y_3^{~}y_4^{~}\right) 
\left(y_{1}^{\prime}y_{2}^{\prime}y_{3}^{\prime}y_{4}^{\prime}\right)}^{n+1} \qquad \\
=\dfrac{1}{4}
\left(
\raisebox{-3em}{\includegraphics[height=6em]{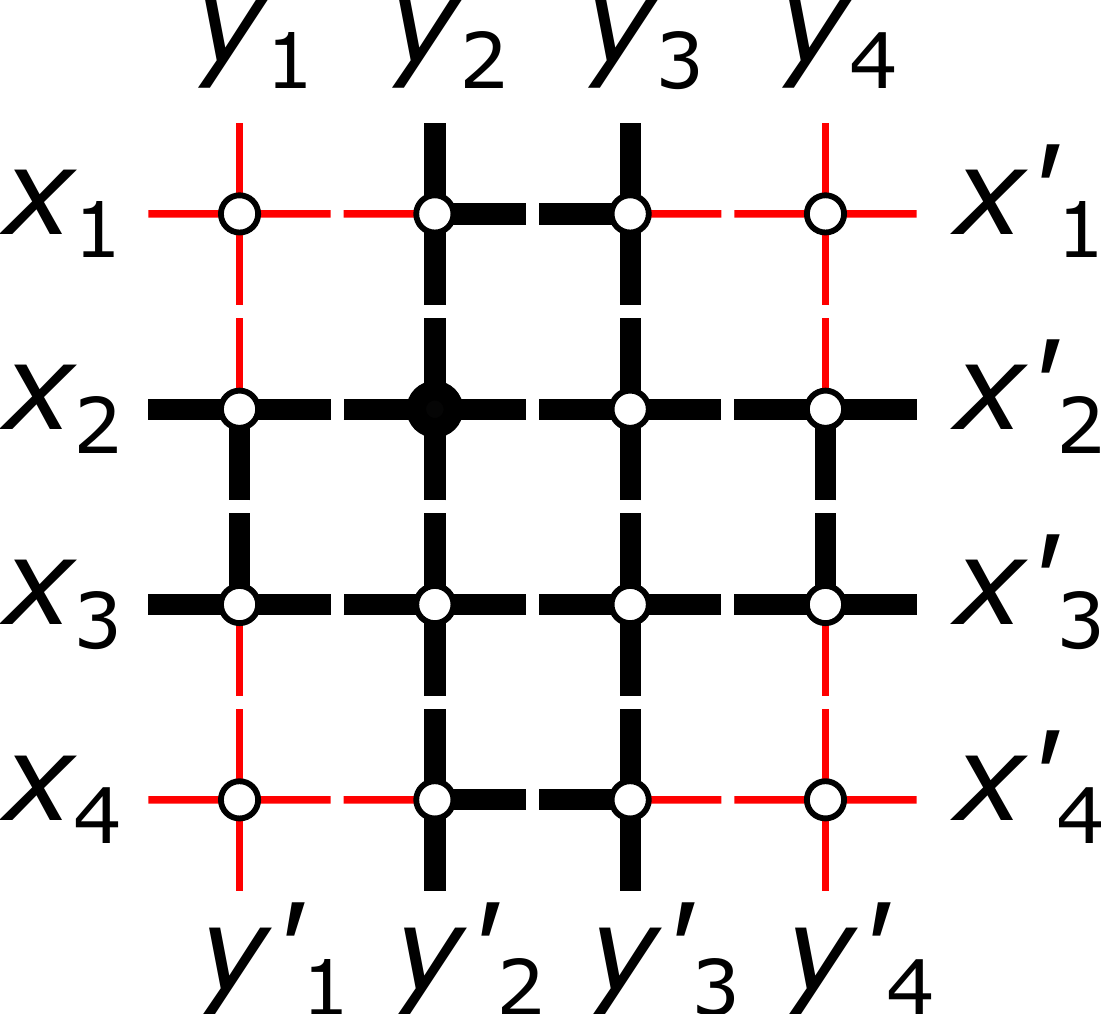}} 
 + 
\raisebox{-3em}{\includegraphics[height=6em]{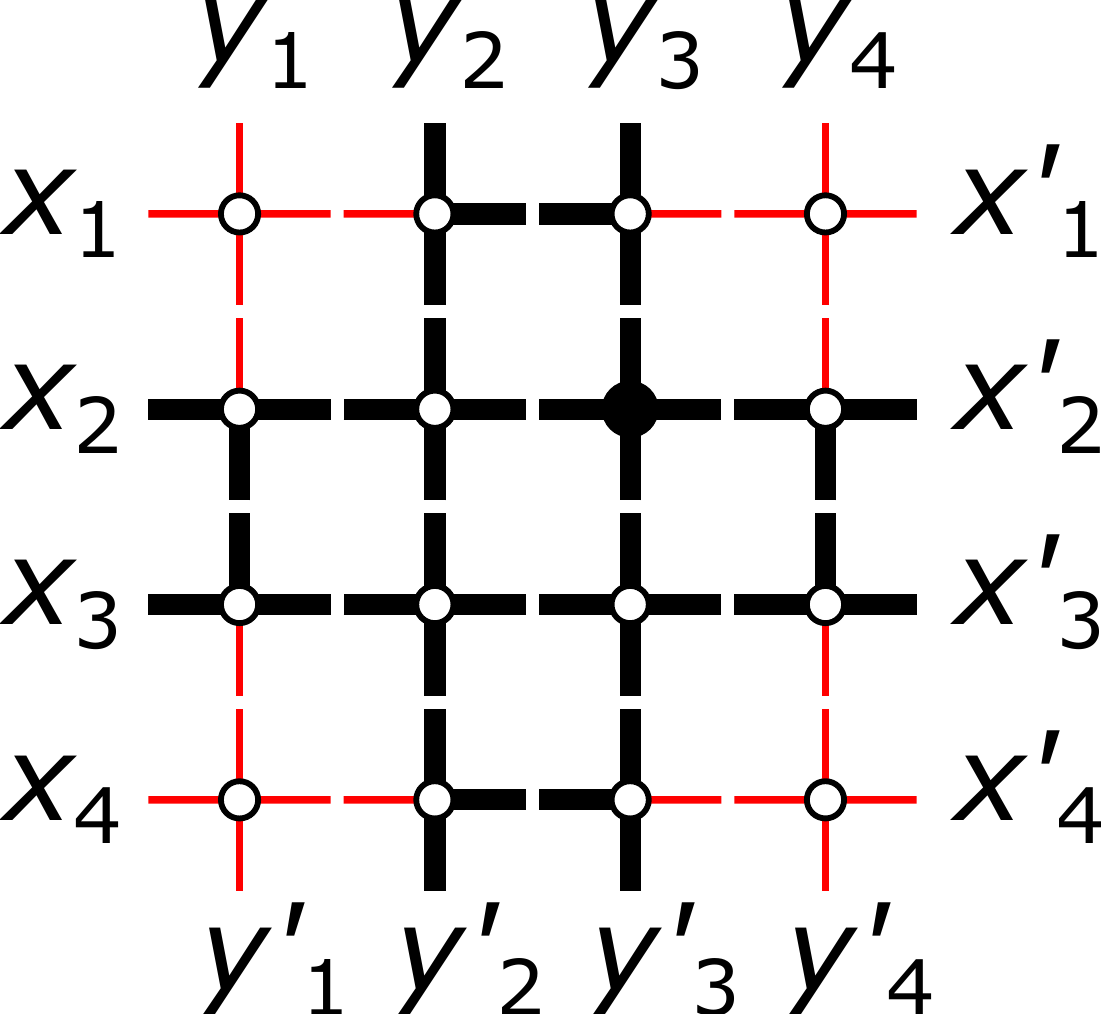}} 
 \right. 
 +  \nonumber
  \\
\left. 
\raisebox{-3em}{\includegraphics[height=6em]{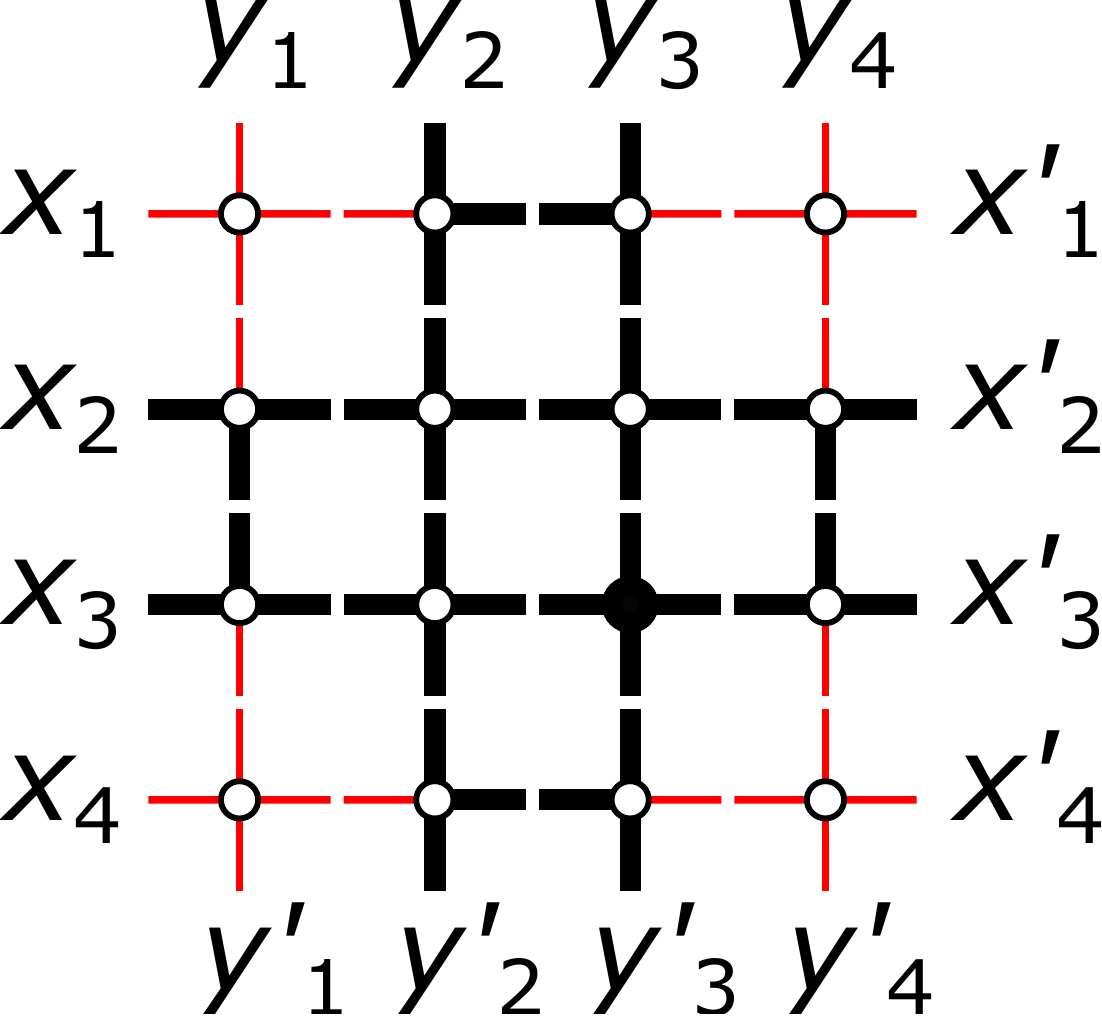}} 
 + 
\raisebox{-3em}{\includegraphics[height=6em]{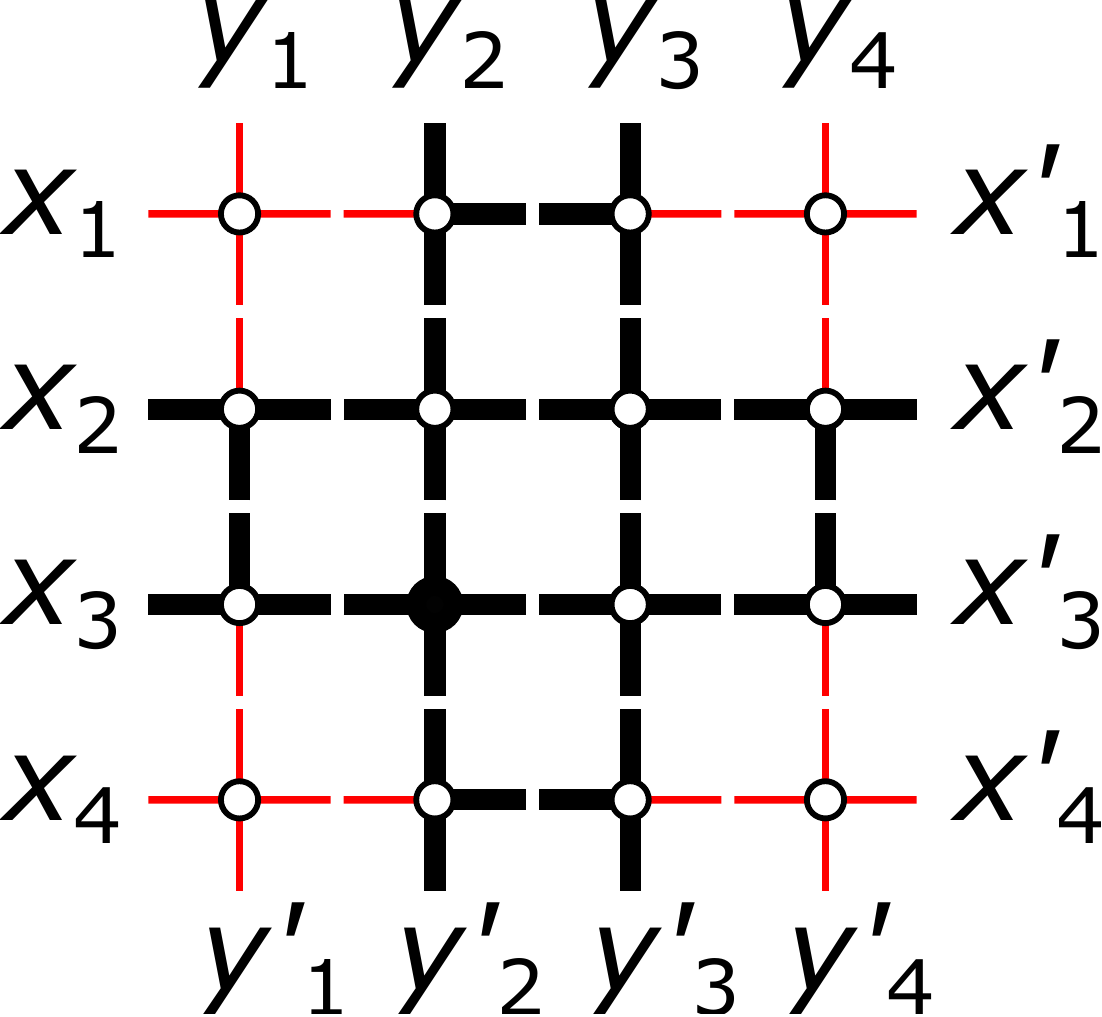}} 
 \right)
 \, ,  \nonumber
\end{eqnarray}
where the full circles represent the inserted impurities.
The explicit form of Eq.~\ref{extension_imp1} is presented in Appendix~\ref{Magnetization_app}.

\noindent
{\it Bond energy}: The bond energy is proportional to the correlation between two nearest-neighbor spins. We start by defining an initial impurity tensor, as we have done for the spontaneous magnetization (see Eq.~{\eqref{mag_imp_init}}). Taking an average over the four bond energies at the central spins, as they correspond to the nearest-neighbor pairs. The first extension step ($n=0$) is performed by taking an average over the four different neighboring pairs of the impurity tensors in the extension pattern ${\cal T}^{[1],n+1}$
\begin{eqnarray} \label{extension_imp2}
{\tilde{\tilde{\cal T}}}_{
\left(x_1^{~}x_2^{~}x_3^{~}x_4^{~}\right) 
\left(x_{1}^{\prime}x_{2}^{\prime}x_{3}^{\prime}x_{4}^{\prime}\right) 
\left(y_1^{~}y_2^{~}y_3^{~}y_4^{~}\right) 
\left(y_{1}^{\prime}y_{2}^{\prime}y_{3}^{\prime}y_{4}^{\prime}\right)}^{n=1} \qquad \\
=\dfrac{1}{4}
\left(
\raisebox{-3em}{\includegraphics[height=6em]{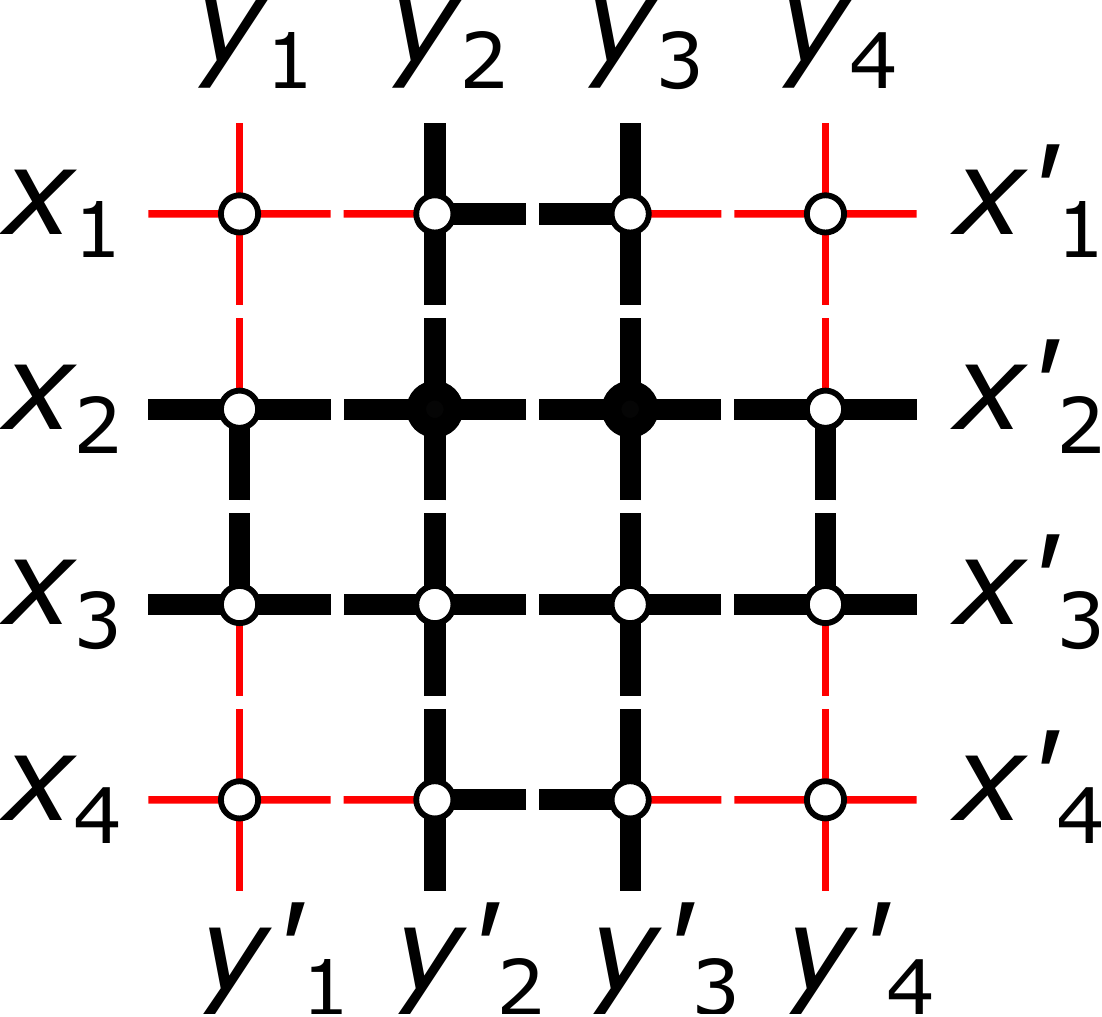}} 
 + 
\raisebox{-3em}{\includegraphics[height=6em]{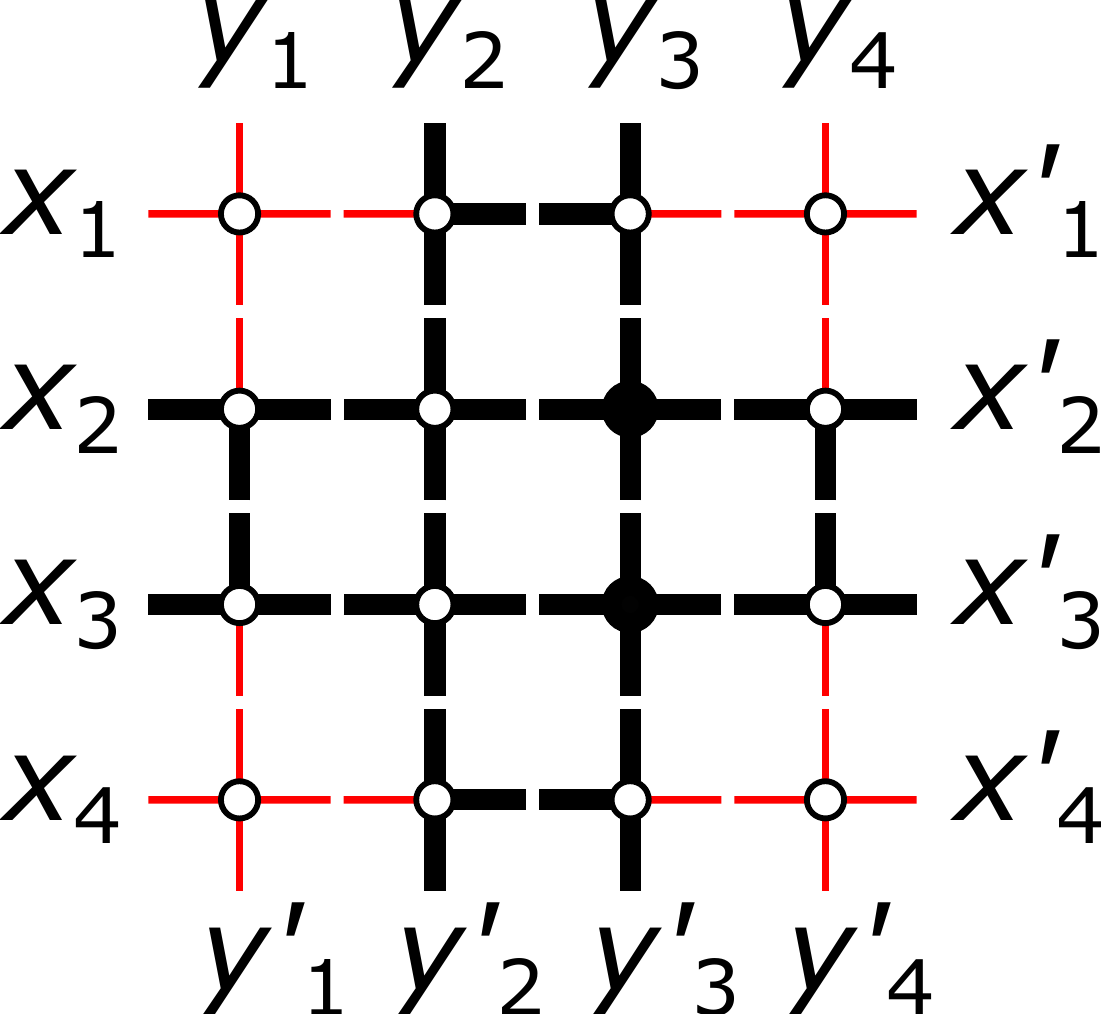}} 
 \right. 
 +  \nonumber
  \\
\left. 
\raisebox{-3em}{\includegraphics[height=6em]{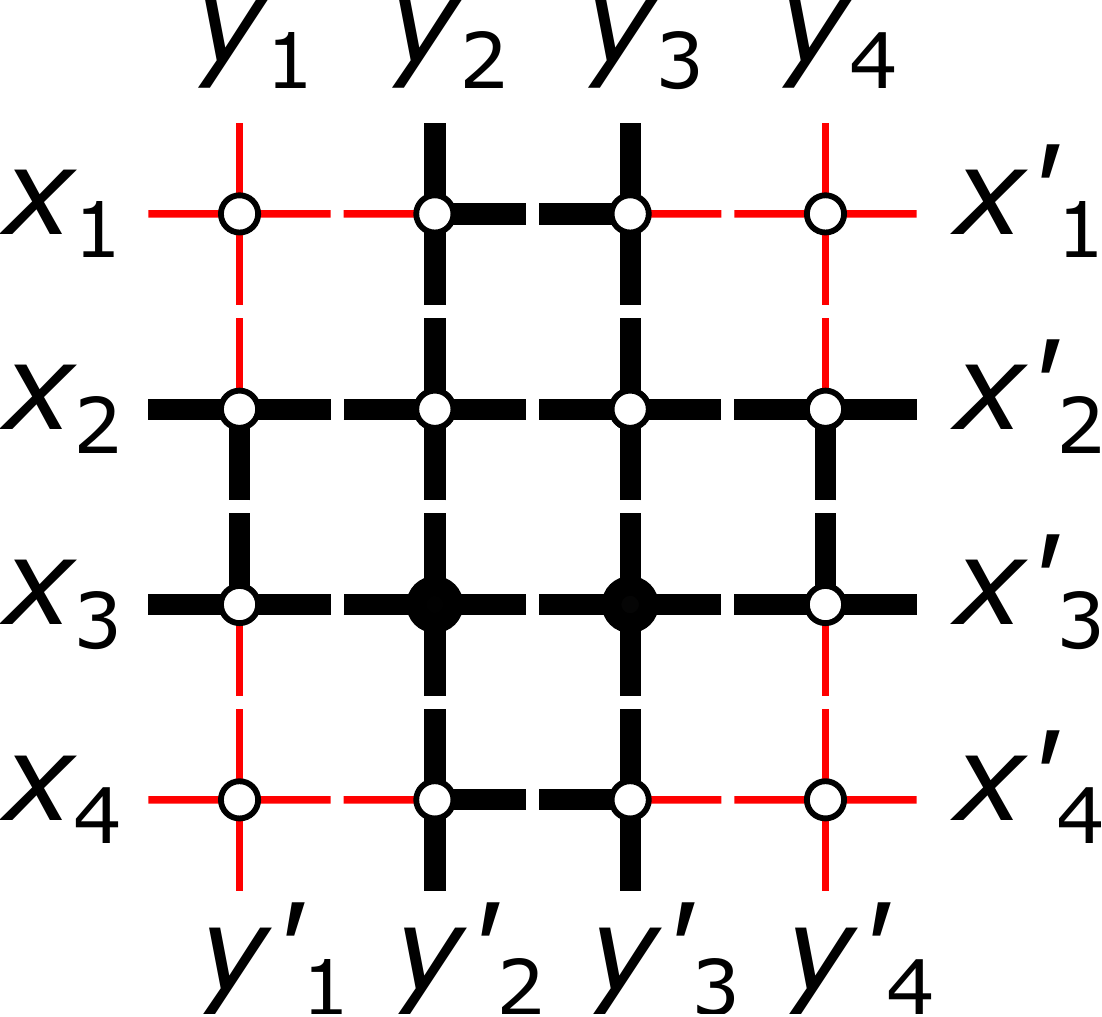}} 
 + 
\raisebox{-3em}{\includegraphics[height=6em]{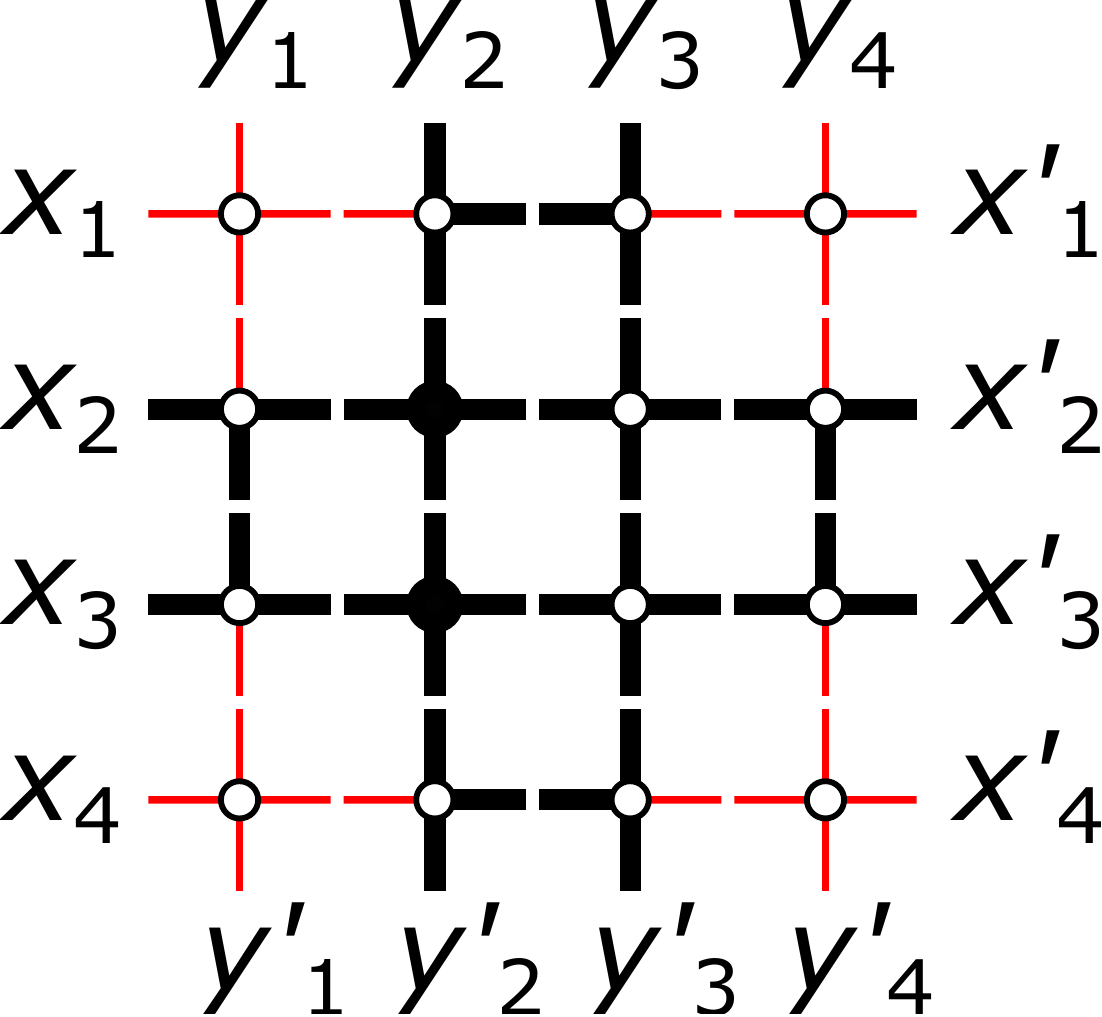}} 
 \right)
 \, ,  \nonumber
\end{eqnarray}
When $n>0$, we extend the bond-energy impurity tensor according to Eq.~\eqref{extension_imp1}. 
The explicit form of Eq.~\ref{extension_imp2} is presented in Appendix~\ref{Bond_energy_app}.

\section{Numerical Results}

We are interested in the critical behavior of the \mbox{$J_1\text{-}J_2$} fractal lattice in the regime between the {\it pure} fractal lattice (i.~e. when $J_1=1$ and $J_2=0$) and the regular square lattice (i.~e. when $J_{1}^{} = J_{2}^{} = 1$). From now on, we will set the fractal coupling to $J_1=1$, only changing $0 \leq J_2 \leq 1$.

\begin{figure}
\includegraphics[width=0.45 \textwidth]{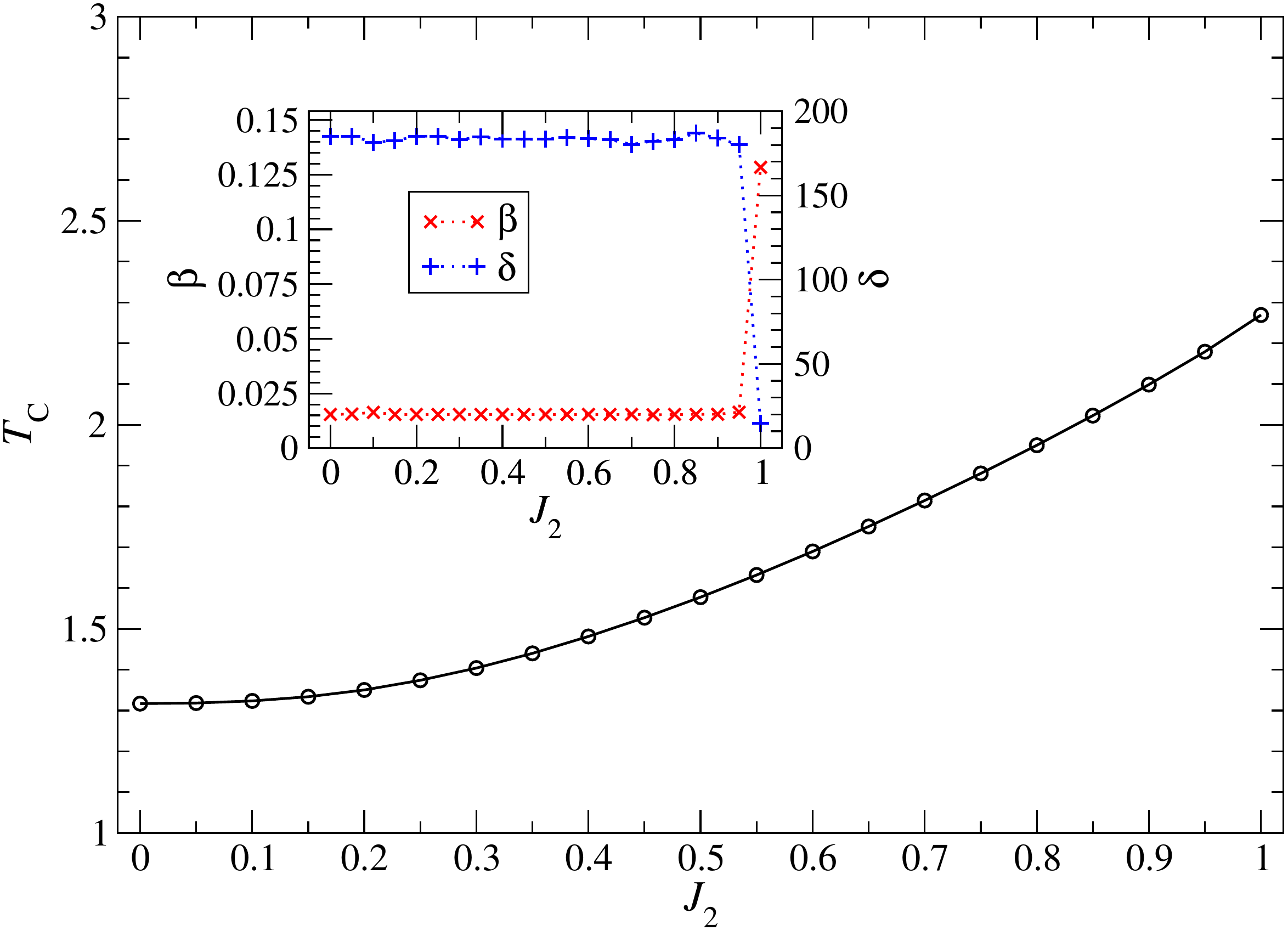} 
\caption{ (Color online) \label{tc_fig}
Critical temperature $T_{\rm c}^{~}$ with respect to $J_2$ (for $J_1 = 1$, $D=34$). Inset: The  $J_2$ dependence of the critical exponents $\beta$ (red) and $\delta$ (blue).
}
\end{figure}

Let us first analyze the spontaneous magnetization (see Fig.~\ref{tc_fig}), where the phase transition (critical) temperature $T_{\rm c}^{~}$ continuously increases as $J_2$ increases. The power-law decay of the spontaneous magnetization, obtained from the impurity tensor $\tilde{\cal T}^{n}$ at $T\leq T_{\rm c}^{~}$ and the external magnetic field $h=0$, below the critical temperature follows the scaling
\begin{equation}
\langle\tilde{\cal T}^{n}\rangle \propto {(T_{\rm c}^{~} - T)}^{\beta}.
\end{equation}
It is important to stress the fact that the critical exponent $\beta$ does not significantly vary within the entire interval of $0 \leq J_2 \lesssim 1$ (see inset of Fig.~\ref{tc_fig}). It remains almost identical to the case of the {\it pure} fractal lattice ($J_2=0$) where $\beta \approx 0.015$ until the square lattice recovers ($J_2=1$), where the exponent suddenly jumps to the expected value (exact value is $\beta_{\rm square} = 1/8 = 0.125$).

Likewise, the other critical exponent $\delta$ associated with the induced magnetization at $T = T_{\rm c}^{~}$ and a nonzero external magnetic field $0 \leq h \lesssim 10^{-7}$
\begin{equation}
\left.\langle\tilde{\cal T}^{n}\rangle \right\rvert_{T = T_{\rm c}^{~}} \propto {h}^{1/\delta} \, ,
\end{equation}
does not significantly vary within $0 \leq J_2 \lesssim 1$ (see inset of Fig.~\ref{tc_fig}). 
The value of $\delta$ does not change from the pure fractal case $\delta \approx 185$ when $J_2 = 0$ until the square lattice recovers ($J_2=1$), where $\delta$ jumps to the exact value known for the square lattice $\delta_{\rm square} = 15$.

In the case of the {\it pure} fractal lattice ($J_2^{~}=0$), 
it is sufficient to employ a moderate value of the bond dimension $D$. 
Our numerical results do not change for $D\geq16$. 
Analyzing the spontaneous magnetization, we found the critical temperature $T_{\rm c}^{~} \approx 1.31695$ 
and the critical exponent $\beta \approx 0.0153951$, which does not change for $D=(16, 18, 20, 22, 24)$). 

When comparing the current results with the previous study~\cite{APS}, 
the critical temperature is almost identical (compare to $T_{\rm c}^{~} = 1.31717$ (with $D=32$), 
which yields the relative difference of $\sim 0.02\%$. 
On the other hand, the magnetic exponent exhibits a larger difference at $D=32$, where $\beta = 0.01388$, thus yielding the relative difference of $\sim 10\%$.

\begin{figure} 
\includegraphics[width=0.48 \textwidth]{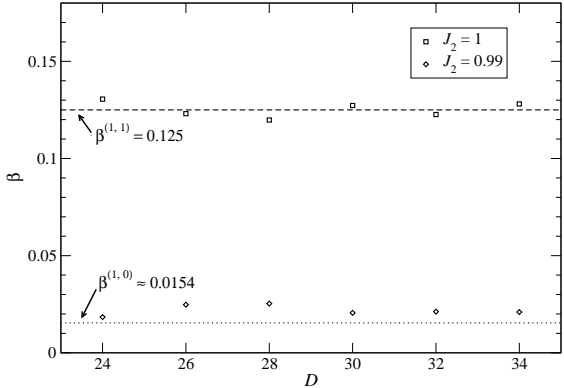}
\caption{ \label{beta_vs_D}
Critical exponent $\beta$ as a function of the bond dimension $D$. 
The numerical results obtained for $J_2=1$ remain close to the exact value $\beta=0.125$ (dashed line).
If setting $J_2=0.99$, the numerical results remain close to the numerical value obtained for the {\it pure} fractal case associated with $J_2=0$, where $\beta \approx 0.0154$ (dotted line).
}
\end{figure}

For $J_2 = 0.99$, we got $T_{\rm c}^{~} \approx 2.24964$ and $\beta \approx 0.021$ (with $D=34$). 

For $J_2 = 1$ (being the regular square lattice), we got $T_{\rm c}^{~} \approx 2.26919$ and $\beta \approx 0.128$ ($D=34$) in full agreement with the exact solution $T_{\rm c}^{~} = 2/\ln\left(1 + \sqrt{2}\right) \approx 2.26919$ and $\beta = 1/8 = 0.125$. 
To show the bond-dimension $D$ dependence, we plot the $\beta$ exponent for $J_2=0.99$ and $J_2=1$ in Fig.~\ref{beta_vs_D}.

For $J_2 = 1.01$ (i.~e., {\it inverse} fractal lattice), we got $T_{\rm c}^{~} \approx 2.28875$ and $\beta \approx 0.5$ ($D=34$). 
The critical temperature $T_{\rm c}^{~}$ continues to rise, as we further increase $J_2>1$.
However, the critical exponent $\beta$ seems to be stabilized around the (mean-field) value $\beta \approx 0.5$ in the {\it inverse} fractal regime when $J_2>1$ (not shown, but confirmed up to $J_2=1.1$ at $D=32$).

If studying the magnetic-field response, we introduce a small $h$ into the system at the critical point. 
For the regular square lattice ($J_2=1$), the HOTRG method results in the critical exponent $\delta$ with the relative error of less than one percent (with $D = 34$). 
For the {\it pure} fractal lattice, we obtained $\delta \approx 185$ (with $D=34$). 
Comparing them with the previous results~\cite{APS}, we found a relative difference of $\sim 11\%$ (the previous study yielded $\delta \approx 206$ at $D=12$). 
Interestingly, close to the regular square lattice from the {\it pure} fractal side ($J_2 = 0.99$), we obtained $\delta \approx 129$ (with $D=34$), whereas, from the {\it inverse} fractal side ($J_2 = 1.01$), we found $\delta \approx 5.1$ (with $D=34$). 

\begin{figure}
\includegraphics[width=0.49 \textwidth]{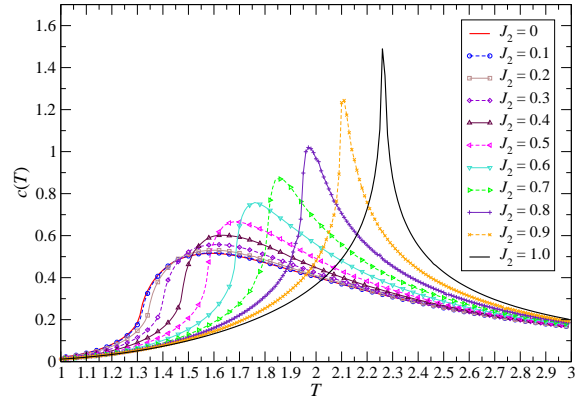}
\caption{ \label{CV1}
Temperature dependence of the specific heat $c(T)$ for various $J_2$ at $D=16$. Notice a weak divergence of $c(T)$ for $J_2<1$ in the inflection point, which refers to the correct phase-transition temperature~\cite{2dising,carpet}.
}
\end{figure}

\begin{figure}
\includegraphics[width=0.49 \textwidth]{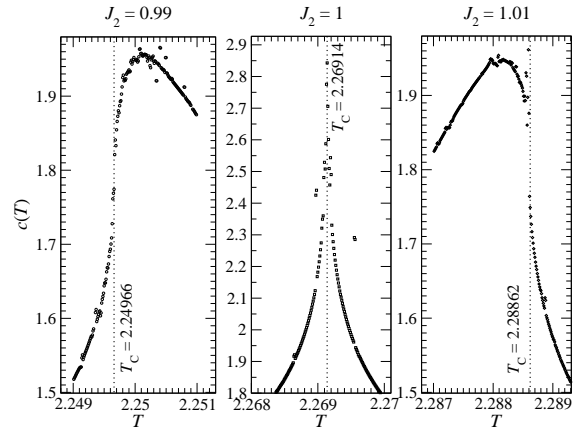}
\caption{ \label{CV2}
The specific heat $c(T)$ around phase transition $T_{\rm c}^{~}$ if $J_2=0.99$, $J_2=1$ and $J_2=1.01$ (at $D=24$).
}
\end{figure}

In order to observe phase transition, we evaluate the specific heat $c(T) = \frac{\rm d}{{\rm d}T}u(T)$ by derivating the bond energy $u(T) = \langle {\tilde{\tilde{\cal T}}}^{n} \rangle$ in accord with Eq.~\eqref{extension_imp2}.
The specific heat $c(T)$ does not diverge at any value of $J_2 < 1$, see Fig.~\ref{CV1}. 
We observe the divergence only in the case of the regular square lattice, i.e., when $J_2 = 1$. If $0\leq J_2 < 1$, the maximum of the specific heat $c(T)$ does not correspond to the critical point. Instead, we have confirmed and numerically verified (see Refs.~\onlinecite{2dising} and \onlinecite{APS}) that the maximum of the first numerical derivative with respect to $T$ corresponds to the critical point for all values of $J_2 < 1$, i.e.,
\begin{equation}
	T_{\rm c}^{~} = \max_T \left\{\frac{\rm d}{{\rm d}T}c_{}(T)\right\}.
\end{equation}
We have also explored the vicinity of $J_2=1$ and considered the cases when $J_2=0.99$ and $J_2=1.01$. There, the singularity of $c(T)$ at $T_{\rm c}^{~}$ appears only if $J_2=1$, as depicted in Fig.~\ref{CV2}. 

\begin{figure}
\includegraphics[width=0.48 \textwidth]{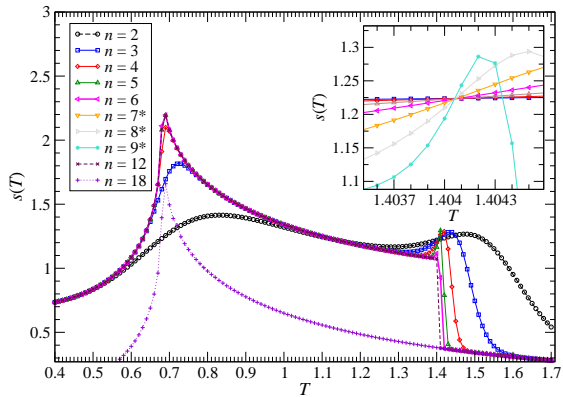}
\caption{ \label{Entropy1} 
Entanglement entropy $s(T)$ for $J_2 = 0.3$ ($D=16$) measured at various iteration steps $n$. The left peak in $s(T)$ is converged after six iterations ($n \geq 6$) and is located at $T = J_2 T_{\rm c}^{(J_2=1)} \approx 0.68$. The entanglement entropy is invariant at $3 \leq n \leq 9$ resulting in the critical point of the fractal at $T_{\rm c}^{(J_2=0.3)} \approx 1.40405$. The values of $s(T)$ denoted by stars are not shown in the main plot for better visibility (for $n=7,8,9$). Inset: the zoomed-in view, where the entropy has the fixed point at $T_{\rm c}^{(J_2=0.3)}$.
}
\end{figure}

\begin{figure}
\includegraphics[width=0.48 \textwidth]{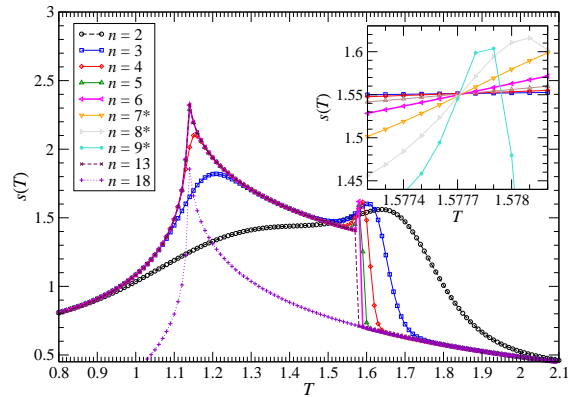}
\caption{ \label{Entropy2} 
Entanglement entropy $s(T)$ for $J_2 = 0.5$ ($D=16$). The left peak in $s(T)$ appears at $T = J_2 T_{\rm c}^{(J_2=1)} \approx 0.68$. The entanglement entropy is invariant at $T_{\rm c}^{(J_2=0.5)} \approx 1.5777$. Inset: the detail of the entropy fixed point. 
}
\end{figure}

In order to determine the phase phase transition correctly, we analyze entanglement entropy $s(T)$ introduced in Eq.~\eqref{entanglement}, which usually achieves full numerical convergence after $n \approx 18$ iteration steps. We select two vales $J_2 = 0.3$ and $J_2 = 0.5$ (at $D=16$) to observe how the iteration steps $n$ affect the entanglement entropy $s(T)$, as plotted in Fig.~\ref{Entropy1} and Fig.~\ref{Entropy2}, respectively.
Surprisingly, the converged entanglement entropy exhibits a divergence at temperatures $T\approx 0.68$ ($J_2 = 0.3$) and $T\approx 1.13$ ($J_2 = 0.5$) which substantially differs from the expected fractal critical temperatures $T_{\rm c}^{~} \approx 1.40405$ and $T_{\rm c}^{~} \approx 1.5777$, respectively. We have numerically confirmed for additional values of $0<J_2<1$ that the sharp maxima of the converged entanglement entropy $n>18$, typically associated with the phase transition, occurs at
\begin{equation}
T = J_2 T_{\rm c}^{(J_2=1)} =  \frac{2J_2}{\ln\left(1 + \sqrt{2}\right)}\, ,
\label{J2Tc}
\end{equation}
which corresponds to the critical temperature of the regular-square lattice Ising model $ T_{\rm c}^{(J_2=1)}$ multiplied by the coupling $J_2$. The entanglement entropy $s(T)$ at $J_2 = 0.5$ exhibits an interesting behavior, see Fig.~\ref{Entropy2}, after six iterations $s(T)$ yields two peaks: the left peak around $T\approx 1.13$ and the right peak around the correct fractal critical temperature $T\approx 1.5777$ at $J_2 = 0.5$. If zooming-in around the fractal critical temperature, the entanglement entropy becomes invariant in between $n=3$ and $n=9$. Such fixed-point behavior captures the correct phase transition of the fractal. 


\section{Conclusions and Discussions}

We have investigated the Ising model on a continuous family of planar self-similar lattices with two types of coupling strengths, $J_1$ and $J_2$. 
For this purpose, we have developed a modified HOTRG technique with multiple types of local tensors; each one being extended with specific coarse-graining patterns. 
We focused on the way of how critical behavior of the Ising model on the self-similar lattices changes while transforming the lattice from the fractal pattern into the regular square lattice by the continuous change of the value of $J_2$ from zero to one (if fixing $J_1=1$).
The critical temperature $T_{\rm c}^{~}$, as captured by the local order parameter (the spontaneous magnetization) as well as the local bond energy (the internal energy), grows continuously from $T_{\rm c}^{(J_2=0)} \approx 1.317$ (correct value for the fractal lattice as shown in Ref.~\cite{2dising, APS}) to $T_{\rm c}^{(J_2=1)} \approx 2.269$ (known exactly). 
Having analyzed the magnetic critical exponent $\beta$ when tuning $0 \leq J_2 \leq 1$, we did not observe a continuous change of the exponent in the interval $0.015 \lesssim \beta \leq \frac{1}{8}$, as one would have naturally expected. Instead, we determined almost a constant value of $\beta \approx 0.015$ on the entire interval $0 \leq J_2 < 1$ followed by a significant discontinuous jump if $J_2=1$, where $\beta = \frac{1}{8}$.
The other magnetic critical exponent $\delta$ exhibited similar singular behavior. We measured a constant value of $\delta \approx 185$ on the interval $0 \leq J_2 < 1$ followed by a significant discontinuous jump if $J_2=1$, where $\delta = 15$.

The specific heat $c(T)$ also followed qualitatively similar behavior as the magnetic exponent $\beta$. A sharp singularity appeared at $J_2=1$ only, whereas broadened maxima of $c(T)$ did not coincide with the correct $T_{\rm c}^{~}$ determined from the spontaneous magnetization in Fig.~\ref{tc_fig}. Instead, the sharp peaks of $\frac{\rm d}{{\rm d}T}c(T)$ referred to the correct $T_{\rm c}^{~}$ (in accord with Ref.~\onlinecite{2dising}).

To address the question of why we observed the discontinuity in $\beta$ when $J_2=1$ (including analogous behavior in the specific heat), we relate the answer to the fact that the $J_1$-$J_2$ model on the lattice becomes fully translationally invariant only if $J_1=J_2$. 
Otherwise, the $J_1$-$J_2$ fractals possess a weaker type of symmetry, i.~e. the scale invariance.  
Based on our numerical observations, we conjecture that there are three classes of the behavior of the $J_1$-$J_2$ fractals: (i) fractal-like when $J_1<J_2$ (with $\beta \approx 0.02$), (ii) regular square lattice when $J_1=J_2$ (with $\beta = 1/8$), and (iii) {\it inverse} fractal when $J_1>J_2$ (with the mean-field value of $\beta \approx 0.5$).

The entanglement entropy $s(T)$ calculated from the singular values obtained as a byproduct of the SVD during the renormalization of the tensors captures the global behavior of the system.
The entanglement entropy after around $n=18$ iteration steps corresponds to the regular square lattice, whose critical point uniformly scales with the prefactor $J_2$ and information on the fractal structure is suppressed. 
However, when the number of the iteration step in HOTRG does not reach the full numerical convergence with spontaneously broken symmetry, $s(T)$ can capture two phase transitions: (i) the one occurring inside the fractal structure which is in an agreement with the local quantities such as the spontaneous magnetization and (ii) the phase transition taking place on the homogeneous part at the critical temperature $T = J_2 T_{\rm c}^{(J_2=1)}$, cf. Eq.~\eqref{J2Tc}.
The former case corresponds to the fixed-point-like behavior of the entanglement entropy $s(T)$ with respect to the iteration steps $n$.

Another interesting question arises: How to think about the dimensionality of the lattice, between the fractal and the regular square lattice. 
We propose that it might be meaningful to define an appropriate dimension, in this case, being a combination of both coupling strengths $J_1$ and $J_2$. 
A generalized Hausdorff dimension might have been defined
\begin{equation}
	d_1 = \dfrac{\ln (12 J_1 + 4 J_2)}{\ln 4} \, .
\end{equation}
If considering the scaling of the boundary bonds, the other dimension could follow the expression
\begin{equation}
d_2 = 1 + \dfrac{\ln (2 J_1 + 2 J_2)}{\ln 4} \, .
\end{equation}
Both formulas need to be adapted to the case when \mbox{$J_2=0$}. Then, irrespective of $J_1$, we need to recover $d_1=\ln 12 / \ln 4$ and $d_2 = 1.5$ (see Ref.~\onlinecite{2dising}), provided that \mbox{$J_1 \neq 0$}.

The current study opens the door to many exciting directions of the research.
HOTRG can be applied to the study of the quantum Ising model on the \mbox{$J_1\text{-}J_2$} fractals. 
Also, the extension to three dimensions of the correct classical spin model is possible, although computationally very complex and requiring more extensive computational resources.
Moreover, the technique presented in this work is applicable to a variety of non-homogeneous lattices, including the Sierpi\'{n}ski carpet.
In some cases, it might be desirable to employ different optimization schema of the local tensors than used here; however, the basic idea of extending multiple types of the local tensors remains valid. 
Apart from the typical condensed-matter applications, our technique might inspire new data compression approaches, for example, in image processing.

%
%
%
%
%
%
%
%

\begin{acknowledgments}
The support received from the project OPTIQUTE APVV-18-0518, EXSES
APVV-16-0186, Joint Research Project SAS-MOST 108-2112-M-002-020-MY3, and VEGA Grants No. 2/0123/19 is acknowledged.

\end{acknowledgments}

\appendix

\section{Extension Patterns} \label{app_A}

Each of the ten tensor types can be extended by means of a different extension relation (see Table~\ref{table:Fig_2}). 
For instance, the extension formula for the tensor type ${\cal T}_{~}^{[k=1]}$ is shown in Eq.~\eqref{Eq_Crazy}. 

Consider the extension of the tensor ${\cal T}_{~}^{[1]}$ in detail. 
As seen from Table~\ref{table:Fig_2}, to obtain the new tensor ${\cal T}_{~}^{[1], n+1}$, one needs to contract 16 tensors in total (four tensors of type 1, two of type 2, two of type 3, two of type 4, two of type 5, and finally four of type 10) arranged on a $4\times4$ grid according to the specified pattern
\begin{eqnarray}
\label{Eq_Crazy_app}
{\cal T}_{
\left(x_1^{~}x_2^{~}x_3^{~}x_4^{~}\right) 
\left(x_{1}^{\prime}x_{2}^{\prime}x_{3}^{\prime}x_{4}^{\prime}\right) 
\left(y_1^{~}y_2^{~}y_3^{~}y_4^{~}\right) 
\left(y_{1}^{\prime}y_{2}^{\prime}y_{3}^{\prime}y_{4}^{\prime}\right)}^{[1], {n+1}} = \\
=\sum_{\substack{a b c d e f g h i \\ j k l m n^{\prime} o p q \\ r s t u v w x}}^{~}
{{{{\cal T}^{[10], n}_{x_1^{~} a^{~}_{~} y_1^{~} d^{~}_{~}}  
{\cal T}^{[5], n}_{a^{~}_{~} b^{~}_{~} y_{2}^{~} e^{~}_{~}}
{\cal T}^{[3], n}_{b^{~}_{~} c^{~}_{~} y_{3}^{~} f^{~}_{~}} 
{\cal T}^{[10], n}_{c^{~}_{~} x_{1}^{\prime} y_{4}^{~} g^{~}_{~}} } \atop
{{\cal T}^{[2], n}_{x_{2}^{~} h^{~}_{~} d^{~}_{~} k^{~}_{~}}
 {\cal T}^{[1], n}_{h^{~}_{~} i^{~}_{~} e^{~}_{~} l^{~}_{~}}
{\cal T}^{[1], n}_{i^{~}_{~} j^{~}_{~} f^{~}_{~} m^{~}_{~}} 
{\cal T}^{[2], n}_{j^{~}_{~} x_{2}^{\prime} g_{~}^{~} n^{\prime}_{~}}}} \atop
{{{\cal T}^{[4], n}_{x_{3}^{~} o_{~}^{~} k_{~}^{~} r_{~}^{~}}  
  {\cal T}^{[1], n}_{o_{~}^{~} p_{~}^{~} l_{~}^{~} s_{~}^{~}} 
  {\cal T}^{[1], n}_{p_{~}^{~} q_{~}^{~} m_{~}^{~} t_{~}^{~}} 
  {\cal T}^{[4], n}_{q_{~}^{~} x_3^{\prime} n_{~}^{\prime} u_{~}^{~}} } \atop
{{\cal T}^{[10], n}_{x_4^{~} v_{~}^{~} r_{~}^{~} y_1^{\prime}}
 {\cal T}^{[5], n}_{v_{~}^{~} w_{~}^{~} s_{~}^{~} y_{2}^{\prime}}
 {\cal T}^{[3], n}_{w_{~}^{~} x_{~}^{~} t_{~}^{~} y_3^{\prime}}
 {\cal T}^{[10], n}_{x_{~}^{~} x_4^{\prime} u_{~}^{~} y_4^{\prime}}}}} 
 \, ,  \nonumber
\end{eqnarray}

Here, we list all ten extension relations in an abbreviated form where we omitted all the tensor indices except for the tensor-type index (in square brackets) and the iteration step $n$ (the omitted tensor indices are identical to those in Eq.~\eqref{Eq_Crazy_app} in all the remaining formulas). 
The new tensors ${\cal T}_{~}^{[k], n+1}$ have been obtained from the preceding iteration step, out of the tensors ${\cal T}_{~}^{[k], n}$, where $k=1,2, \dots ,10$. 
Each extension relation specifies the pattern of the 16 previously prepared tensors ${\cal T}_{~}^{n}$ on a $4\times4$ grid (on the right-hand side of the formulas below), which are needed to obtain the extended ${\cal T}_{~}^{n+1}$ tensors of each type (on the left-hand side). 

\begin{eqnarray}
{T}_{{~}_{~}^{~} {~}_{~}^{~}}^{[1], {n+1}} &\leftarrow&
\left(
{{{{T}^{[10] ~}_{{~}_{~}^{~} {~}_{~}^{~}}  
{T}^{[5] ~}_{{~}_{~}^{~} {~}_{~}^{~}}
{T}^{[3] ~}_{{~}_{~}^{~} {~}_{~}^{~}} 
{T}^{[10] ~}_{{~}_{~}^{~} {~}_{~}^{~}} } \atop
{{T}^{[2] ~}_{{~}_{~}^{~} {~}_{~}^{~}}
 {T}^{[1] ~}_{{~}_{~}^{~} {~}_{~}^{~}}
{T}^{[1] ~}_{{~}_{~}^{~} {~}_{~}^{~}} 
{T}^{[2] ~}_{{~}_{~}^{~} {~}_{~}^{~}}}} \atop
{{{T}^{[4] ~}_{{~}_{~}^{~} {~}_{~}^{~}}  
  {T}^{[1] ~}_{{~}_{~}^{~} {~}_{~}^{~}} 
  {T}^{[1] ~}_{{~}_{~}^{~} {~}_{~}^{~}} 
  {T}^{[4] ~}_{{~}_{~}^{~} {~}_{~}^{~}} } \atop
{{T}^{[10] ~}_{{~}_{~}^{~} {~}_{~}^{~}}
 {T}^{[5] ~}_{{~}_{~}^{~} {~}_{~}^{~}}
 {T}^{[3] ~}_{{~}_{~}^{~} {~}_{~}^{~}}
 {T}^{[10] ~}_{{~}_{~}^{~} {~}_{~}^{~}}}}} \right)^{, n} \, ,  \nonumber 
 \\
 {T}_{{~}_{~}^{~} {~}_{~}^{~}}^{[2], {n+1}} &\leftarrow&
\left(
{{{{T}^{[10] ~}_{{~}_{~}^{~} {~}_{~}^{~}}  
{T}^{[9] ~}_{{~}_{~}^{~} {~}_{~}^{~}}
{T}^{[6] ~}_{{~}_{~}^{~} {~}_{~}^{~}} 
{T}^{[10] ~}_{{~}_{~}^{~} {~}_{~}^{~}} } \atop
{{T}^{[2] ~}_{{~}_{~}^{~} {~}_{~}^{~}}
 {T}^{[1] ~}_{{~}_{~}^{~} {~}_{~}^{~}}
{T}^{[1] ~}_{{~}_{~}^{~} {~}_{~}^{~}} 
{T}^{[2] ~}_{{~}_{~}^{~} {~}_{~}^{~}}}} \atop
{{{T}^{[4] ~}_{{~}_{~}^{~} {~}_{~}^{~}}  
  {T}^{[1] ~}_{{~}_{~}^{~} {~}_{~}^{~}} 
  {T}^{[1] ~}_{{~}_{~}^{~} {~}_{~}^{~}} 
  {T}^{[4] ~}_{{~}_{~}^{~} {~}_{~}^{~}} } \atop
{{T}^{[10] ~}_{{~}_{~}^{~} {~}_{~}^{~}}
 {T}^{[5] ~}_{{~}_{~}^{~} {~}_{~}^{~}}
 {T}^{[3] ~}_{{~}_{~}^{~} {~}_{~}^{~}}
 {T}^{[10] ~}_{{~}_{~}^{~} {~}_{~}^{~}}}}} \right)^{, n} \, ,  \nonumber
  \\
 {T}_{{~}_{~}^{~} {~}_{~}^{~}}^{[3], {n+1}} &\leftarrow&
\left(
{{{{T}^{[10] ~}_{{~}_{~}^{~} {~}_{~}^{~}}  
{T}^{[5] ~}_{{~}_{~}^{~} {~}_{~}^{~}}
{T}^{[3] ~}_{{~}_{~}^{~} {~}_{~}^{~}} 
{T}^{[10] ~}_{{~}_{~}^{~} {~}_{~}^{~}} } \atop
{{T}^{[2] ~}_{{~}_{~}^{~} {~}_{~}^{~}}
 {T}^{[1] ~}_{{~}_{~}^{~} {~}_{~}^{~}}
{T}^{[1] ~}_{{~}_{~}^{~} {~}_{~}^{~}} 
{T}^{[6] ~}_{{~}_{~}^{~} {~}_{~}^{~}}}} \atop
{{{T}^{[4] ~}_{{~}_{~}^{~} {~}_{~}^{~}}  
  {T}^{[1] ~}_{{~}_{~}^{~} {~}_{~}^{~}} 
  {T}^{[1] ~}_{{~}_{~}^{~} {~}_{~}^{~}} 
  {T}^{[7] ~}_{{~}_{~}^{~} {~}_{~}^{~}} } \atop
{{T}^{[10] ~}_{{~}_{~}^{~} {~}_{~}^{~}}
 {T}^{[5] ~}_{{~}_{~}^{~} {~}_{~}^{~}}
 {T}^{[3] ~}_{{~}_{~}^{~} {~}_{~}^{~}}
 {T}^{[10] ~}_{{~}_{~}^{~} {~}_{~}^{~}}}}} \right)^{, n} \, ,  \nonumber
 \\
 {T}_{{~}_{~}^{~} {~}_{~}^{~}}^{[4], {n+1}} &\leftarrow&
\left(
{{{{T}^{[10] ~}_{{~}_{~}^{~} {~}_{~}^{~}}  
{T}^{[5] ~}_{{~}_{~}^{~} {~}_{~}^{~}}
{T}^{[3] ~}_{{~}_{~}^{~} {~}_{~}^{~}} 
{T}^{[10] ~}_{{~}_{~}^{~} {~}_{~}^{~}} } \atop
{{T}^{[2] ~}_{{~}_{~}^{~} {~}_{~}^{~}}
 {T}^{[1] ~}_{{~}_{~}^{~} {~}_{~}^{~}}
{T}^{[1] ~}_{{~}_{~}^{~} {~}_{~}^{~}} 
{T}^{[2] ~}_{{~}_{~}^{~} {~}_{~}^{~}}}} \atop
{{{T}^{[4] ~}_{{~}_{~}^{~} {~}_{~}^{~}}  
  {T}^{[1] ~}_{{~}_{~}^{~} {~}_{~}^{~}} 
  {T}^{[1] ~}_{{~}_{~}^{~} {~}_{~}^{~}} 
  {T}^{[4] ~}_{{~}_{~}^{~} {~}_{~}^{~}} } \atop
{{T}^{[10] ~}_{{~}_{~}^{~} {~}_{~}^{~}}
 {T}^{[8] ~}_{{~}_{~}^{~} {~}_{~}^{~}}
 {T}^{[7] ~}_{{~}_{~}^{~} {~}_{~}^{~}}
 {T}^{[10] ~}_{{~}_{~}^{~} {~}_{~}^{~}}}}} \right)^{, n} \, ,  \nonumber
  \\
 {T}_{{~}_{~}^{~} {~}_{~}^{~}}^{[5], {n+1}} &\leftarrow&
\left(
{{{{T}^{[10] ~}_{{~}_{~}^{~} {~}_{~}^{~}}  
{T}^{[5] ~}_{{~}_{~}^{~} {~}_{~}^{~}}
{T}^{[3] ~}_{{~}_{~}^{~} {~}_{~}^{~}} 
{T}^{[10] ~}_{{~}_{~}^{~} {~}_{~}^{~}} } \atop
{{T}^{[9] ~}_{{~}_{~}^{~} {~}_{~}^{~}}
 {T}^{[1] ~}_{{~}_{~}^{~} {~}_{~}^{~}}
{T}^{[1] ~}_{{~}_{~}^{~} {~}_{~}^{~}} 
{T}^{[2] ~}_{{~}_{~}^{~} {~}_{~}^{~}}}} \atop
{{{T}^{[8] ~}_{{~}_{~}^{~} {~}_{~}^{~}}  
  {T}^{[1] ~}_{{~}_{~}^{~} {~}_{~}^{~}} 
  {T}^{[1] ~}_{{~}_{~}^{~} {~}_{~}^{~}} 
  {T}^{[4] ~}_{{~}_{~}^{~} {~}_{~}^{~}} } \atop
{{T}^{[10] ~}_{{~}_{~}^{~} {~}_{~}^{~}}
 {T}^{[5] ~}_{{~}_{~}^{~} {~}_{~}^{~}}
 {T}^{[3] ~}_{{~}_{~}^{~} {~}_{~}^{~}}
 {T}^{[10] ~}_{{~}_{~}^{~} {~}_{~}^{~}}}}} \right)^{, n} \, ,  \nonumber
  \\
 {T}_{{~}_{~}^{~} {~}_{~}^{~}}^{[6], {n+1}} &\leftarrow&
\left(
{{{{T}^{[10] ~}_{{~}_{~}^{~} {~}_{~}^{~}}  
{T}^{[9] ~}_{{~}_{~}^{~} {~}_{~}^{~}}
{T}^{[6] ~}_{{~}_{~}^{~} {~}_{~}^{~}} 
{T}^{[10] ~}_{{~}_{~}^{~} {~}_{~}^{~}} } \atop
{{T}^{[2] ~}_{{~}_{~}^{~} {~}_{~}^{~}}
 {T}^{[1] ~}_{{~}_{~}^{~} {~}_{~}^{~}}
{T}^{[1] ~}_{{~}_{~}^{~} {~}_{~}^{~}} 
{T}^{[6] ~}_{{~}_{~}^{~} {~}_{~}^{~}}}} \atop
{{{T}^{[4] ~}_{{~}_{~}^{~} {~}_{~}^{~}}  
  {T}^{[1] ~}_{{~}_{~}^{~} {~}_{~}^{~}} 
  {T}^{[1] ~}_{{~}_{~}^{~} {~}_{~}^{~}} 
  {T}^{[7] ~}_{{~}_{~}^{~} {~}_{~}^{~}} } \atop
{{T}^{[10] ~}_{{~}_{~}^{~} {~}_{~}^{~}}
 {T}^{[5] ~}_{{~}_{~}^{~} {~}_{~}^{~}}
 {T}^{[3] ~}_{{~}_{~}^{~} {~}_{~}^{~}}
 {T}^{[10] ~}_{{~}_{~}^{~} {~}_{~}^{~}}}}} \right)^{, n} \, ,  \nonumber
  \\
 {T}_{{~}_{~}^{~} {~}_{~}^{~}}^{[7], {n+1}} &\leftarrow&
\left(
{{{{T}^{[10] ~}_{{~}_{~}^{~} {~}_{~}^{~}}  
{T}^{[5] ~}_{{~}_{~}^{~} {~}_{~}^{~}}
{T}^{[3] ~}_{{~}_{~}^{~} {~}_{~}^{~}} 
{T}^{[10] ~}_{{~}_{~}^{~} {~}_{~}^{~}} } \atop
{{T}^{[2] ~}_{{~}_{~}^{~} {~}_{~}^{~}}
 {T}^{[1] ~}_{{~}_{~}^{~} {~}_{~}^{~}}
{T}^{[1] ~}_{{~}_{~}^{~} {~}_{~}^{~}} 
{T}^{[6] ~}_{{~}_{~}^{~} {~}_{~}^{~}}}} \atop
{{{T}^{[4] ~}_{{~}_{~}^{~} {~}_{~}^{~}}  
  {T}^{[1] ~}_{{~}_{~}^{~} {~}_{~}^{~}} 
  {T}^{[1] ~}_{{~}_{~}^{~} {~}_{~}^{~}} 
  {T}^{[7] ~}_{{~}_{~}^{~} {~}_{~}^{~}} } \atop
{{T}^{[10] ~}_{{~}_{~}^{~} {~}_{~}^{~}}
 {T}^{[8] ~}_{{~}_{~}^{~} {~}_{~}^{~}}
 {T}^{[7] ~}_{{~}_{~}^{~} {~}_{~}^{~}}
 {T}^{[10] ~}_{{~}_{~}^{~} {~}_{~}^{~}}}}} \right)^{, n} \, ,  \nonumber
 \\
 {T}_{{~}_{~}^{~} {~}_{~}^{~}}^{[8], {n+1}} &\leftarrow&
\left(
{{{{T}^{[10] ~}_{{~}_{~}^{~} {~}_{~}^{~}}  
{T}^{[5] ~}_{{~}_{~}^{~} {~}_{~}^{~}}
{T}^{[3] ~}_{{~}_{~}^{~} {~}_{~}^{~}} 
{T}^{[10] ~}_{{~}_{~}^{~} {~}_{~}^{~}} } \atop
{{T}^{[9] ~}_{{~}_{~}^{~} {~}_{~}^{~}}
 {T}^{[1] ~}_{{~}_{~}^{~} {~}_{~}^{~}}
{T}^{[1] ~}_{{~}_{~}^{~} {~}_{~}^{~}} 
{T}^{[2] ~}_{{~}_{~}^{~} {~}_{~}^{~}}}} \atop
{{{T}^{[8] ~}_{{~}_{~}^{~} {~}_{~}^{~}}  
  {T}^{[1] ~}_{{~}_{~}^{~} {~}_{~}^{~}} 
  {T}^{[1] ~}_{{~}_{~}^{~} {~}_{~}^{~}} 
  {T}^{[4] ~}_{{~}_{~}^{~} {~}_{~}^{~}} } \atop
{{T}^{[10] ~}_{{~}_{~}^{~} {~}_{~}^{~}}
 {T}^{[8] ~}_{{~}_{~}^{~} {~}_{~}^{~}}
 {T}^{[7] ~}_{{~}_{~}^{~} {~}_{~}^{~}}
 {T}^{[10] ~}_{{~}_{~}^{~} {~}_{~}^{~}}}}} \right)^{, n} \, ,  \nonumber
  \\
 {T}_{{~}_{~}^{~} {~}_{~}^{~}}^{[9], {n+1}} &\leftarrow&
\left(
{{{{T}^{[10] ~}_{{~}_{~}^{~} {~}_{~}^{~}}  
{T}^{[9] ~}_{{~}_{~}^{~} {~}_{~}^{~}}
{T}^{[6] ~}_{{~}_{~}^{~} {~}_{~}^{~}} 
{T}^{[10] ~}_{{~}_{~}^{~} {~}_{~}^{~}} } \atop
{{T}^{[9] ~}_{{~}_{~}^{~} {~}_{~}^{~}}
 {T}^{[1] ~}_{{~}_{~}^{~} {~}_{~}^{~}}
{T}^{[1] ~}_{{~}_{~}^{~} {~}_{~}^{~}} 
{T}^{[2] ~}_{{~}_{~}^{~} {~}_{~}^{~}}}} \atop
{{{T}^{[8] ~}_{{~}_{~}^{~} {~}_{~}^{~}}  
  {T}^{[1] ~}_{{~}_{~}^{~} {~}_{~}^{~}} 
  {T}^{[1] ~}_{{~}_{~}^{~} {~}_{~}^{~}} 
  {T}^{[4] ~}_{{~}_{~}^{~} {~}_{~}^{~}} } \atop
{{T}^{[10] ~}_{{~}_{~}^{~} {~}_{~}^{~}}
 {T}^{[5] ~}_{{~}_{~}^{~} {~}_{~}^{~}}
 {T}^{[3] ~}_{{~}_{~}^{~} {~}_{~}^{~}}
 {T}^{[10] ~}_{{~}_{~}^{~} {~}_{~}^{~}}}}} \right)^{, n} \, ,  \nonumber 
 \\
  {T}_{{~}_{~}^{~} {~}_{~}^{~}}^{[10], {n+1}} &\leftarrow&
\left(
{{{{T}^{[10] ~}_{{~}_{~}^{~} {~}_{~}^{~}}  
{T}^{[10] ~}_{{~}_{~}^{~} {~}_{~}^{~}}
{T}^{[10] ~}_{{~}_{~}^{~} {~}_{~}^{~}} 
{T}^{[10] ~}_{{~}_{~}^{~} {~}_{~}^{~}} } \atop
{{T}^{[10] ~}_{{~}_{~}^{~} {~}_{~}^{~}}
 {T}^{[10] ~}_{{~}_{~}^{~} {~}_{~}^{~}}
{T}^{[10] ~}_{{~}_{~}^{~} {~}_{~}^{~}} 
{T}^{[10] ~}_{{~}_{~}^{~} {~}_{~}^{~}}}} \atop
{{{T}^{[10] ~}_{{~}_{~}^{~} {~}_{~}^{~}}  
  {T}^{[10] ~}_{{~}_{~}^{~} {~}_{~}^{~}} 
  {T}^{[10] ~}_{{~}_{~}^{~} {~}_{~}^{~}} 
  {T}^{[10] ~}_{{~}_{~}^{~} {~}_{~}^{~}} } \atop
{{T}^{[10] ~}_{{~}_{~}^{~} {~}_{~}^{~}}
 {T}^{[10] ~}_{{~}_{~}^{~} {~}_{~}^{~}}
 {T}^{[10] ~}_{{~}_{~}^{~} {~}_{~}^{~}}
 {T}^{[10] ~}_{{~}_{~}^{~} {~}_{~}^{~}}}}} \right)^{, n} \, .  \nonumber 
\end{eqnarray}

\section{Projection Patterns} \label{app_B}

After the extension process, the external legs are projected by the two sets of the {\it external} projectors $U^{1}_{l}$, $U^{2}_{l}$ ($l=1,2,\dots,6$). 
We use $U^{1}$ when projecting thick legs (in black) and $U^{2}$ when projecting thin legs (in red). 
For example, the projections for the tensor types ${\cal T}_{~}^{[k=1]}$, ${\cal T}_{~}^{[k=2]}$, ${\cal T}_{~}^{[k=6]}$, and ${\cal T}_{~}^{[k=10]}$ are performed as follows
\begin{eqnarray} \label{R1_app} 
{\cal T}^{[1], n+1}_{x_{~}^{~} x_{~}^{\prime} y_{~}^{~} y_{~}^{\prime}} = && \\
=\sum_{\substack{x_1^{~} x_2^{~} x_3^{~} x_4^{~} x_5^{~} x_6^{~} \\ x_1^{\prime} x_2^{\prime} x_3^{\prime} x_4^{\prime} x_5^{\prime} x_6^{\prime} \\ y_1^{~} y_2^{~} y_3^{~} y_4^{~} y_5^{~} y_6^{~} \\ y_1^{\prime} y_2^{\prime} y_3^{\prime} y_4^{\prime} y_5^{\prime} y_6^{\prime}}}^{~}&&
\substack{
{\cal T}_{
\left(x_1^{~}x_2^{~}x_3^{~}x_4^{~}\right) 
\left(x_{1}^{\prime}x_{2}^{\prime}x_{3}^{\prime}x_{4}^{\prime}\right) 
\left(y_1^{~}y_2^{~}y_3^{~}y_4^{~}\right) 
\left(y_{1}^{\prime}y_{2}^{\prime}y_{3}^{\prime}y_{4}^{\prime}\right)}^{[1], {n+1}} \\
U^{1}_{1, (x_1^{~} x_2^{~}) x_5^{~}} 
U^{1}_{2, (x_3^{~} x_4^{~}) x_6^{~}} 
U^{1}_{5, (x_5^{~} x_6^{~}) x_{~}^{~}} \\
U^{1}_{1, (x_1^{\prime} x_2^{\prime}) x_5^{\prime}} 
U^{1}_{2, (x_3^{\prime} x_4^{\prime}) x_6^{\prime}} 
U^{1}_{5, (x_5^{\prime} x_6^{\prime}) x_{~}^{\prime}} \\
U^{1}_{3, (y_1^{~} y_2^{~}) y_5^{~}} 
U^{1}_{4, (y_3^{~} y_4^{~}) y_6^{~}} 
U^{1}_{6, (y_5^{~} y_6^{~}) y_{~}^{~}} \\
U^{1}_{3, (y_1^{\prime} y_2^{\prime}) y_5^{\prime}} 
U^{1}_{4, (y_3^{\prime} y_4^{\prime}) y_6^{\prime}} 
U^{1}_{6, (y_5^{\prime} y_6^{\prime}) y_{~}^{\prime}}
}
\, ,  \nonumber
\end{eqnarray}
\begin{eqnarray} \label{R2_app}
{\cal T}^{[2], n+1}_{x_{~}^{~} x_{~}^{\prime} y_{~}^{~} y_{~}^{\prime}} = && \\
=\sum_{\substack{x_1^{~} x_2^{~} x_3^{~} x_4^{~} x_5^{~} x_6^{~} \\ x_1^{\prime} x_2^{\prime} x_3^{\prime} x_4^{\prime} x_5^{\prime} x_6^{\prime} \\ y_1^{~} y_2^{~} y_3^{~} y_4^{~} y_5^{~} y_6^{~} \\ y_1^{\prime} y_2^{\prime} y_3^{\prime} y_4^{\prime} y_5^{\prime} y_6^{\prime}}}^{~}&&
\substack{
{\cal T}_{
\left(x_1^{~}x_2^{~}x_3^{~}x_4^{~}\right) 
\left(x_{1}^{\prime}x_{2}^{\prime}x_{3}^{\prime}x_{4}^{\prime}\right) 
\left(y_1^{~}y_2^{~}y_3^{~}y_4^{~}\right) 
\left(y_{1}^{\prime}y_{2}^{\prime}y_{3}^{\prime}y_{4}^{\prime}\right)}^{[2], {n+1}} \\
U^{1}_{1, (x_1^{~} x_2^{~}) x_5^{~}} 
U^{1}_{2, (x_3^{~} x_4^{~}) x_6^{~}} 
U^{1}_{5, (x_5^{~} x_6^{~}) x_{~}^{~}} \\
U^{1}_{1, (x_1^{\prime} x_2^{\prime}) x_5^{\prime}} 
U^{1}_{2, (x_3^{\prime} x_4^{\prime}) x_6^{\prime}} 
U^{1}_{5, (x_5^{\prime} x_6^{\prime}) x_{~}^{\prime}} \\
U^{2}_{3, (y_1^{~} y_2^{~}) y_5^{~}} 
U^{2}_{4, (y_3^{~} y_4^{~}) y_6^{~}} 
U^{2}_{6, (y_5^{~} y_6^{~}) y_{~}^{~}} \\
U^{1}_{3, (y_1^{\prime} y_2^{\prime}) y_5^{\prime}} 
U^{1}_{4, (y_3^{\prime} y_4^{\prime}) y_6^{\prime}} 
U^{1}_{6, (y_5^{\prime} y_6^{\prime}) y_{~}^{\prime}}
}
\, ,  \nonumber
\end{eqnarray}
\begin{eqnarray} \label{R3_app}
{\cal T}^{[6], n+1}_{x_{~}^{~} x_{~}^{\prime} y_{~}^{~} y_{~}^{\prime}} = && \\
=\sum_{\substack{x_1^{~} x_2^{~} x_3^{~} x_4^{~} x_5^{~} x_6^{~} \\ x_1^{\prime} x_2^{\prime} x_3^{\prime} x_4^{\prime} x_5^{\prime} x_6^{\prime} \\ y_1^{~} y_2^{~} y_3^{~} y_4^{~} y_5^{~} y_6^{~} \\ y_1^{\prime} y_2^{\prime} y_3^{\prime} y_4^{\prime} y_5^{\prime} y_6^{\prime}}}^{~}&&
\substack{
{\cal T}_{
\left(x_1^{~}x_2^{~}x_3^{~}x_4^{~}\right) 
\left(x_{1}^{\prime}x_{2}^{\prime}x_{3}^{\prime}x_{4}^{\prime}\right) 
\left(y_1^{~}y_2^{~}y_3^{~}y_4^{~}\right) 
\left(y_{1}^{\prime}y_{2}^{\prime}y_{3}^{\prime}y_{4}^{\prime}\right)}^{[6], {n+1}} \\
U^{1}_{1, (x_1^{~} x_2^{~}) x_5^{~}} 
U^{1}_{2, (x_3^{~} x_4^{~}) x_6^{~}} 
U^{1}_{5, (x_5^{~} x_6^{~}) x_{~}^{~}} \\
U^{2}_{1, (x_1^{\prime} x_2^{\prime}) x_5^{\prime}} 
U^{2}_{2, (x_3^{\prime} x_4^{\prime}) x_6^{\prime}} 
U^{2}_{5, (x_5^{\prime} x_6^{\prime}) x_{~}^{\prime}} \\
U^{2}_{3, (y_1^{~} y_2^{~}) y_5^{~}} 
U^{2}_{4, (y_3^{~} y_4^{~}) y_6^{~}} 
U^{2}_{6, (y_5^{~} y_6^{~}) y_{~}^{~}} \\
U^{1}_{3, (y_1^{\prime} y_2^{\prime}) y_5^{\prime}} 
U^{1}_{4, (y_3^{\prime} y_4^{\prime}) y_6^{\prime}} 
U^{1}_{6, (y_5^{\prime} y_6^{\prime}) y_{~}^{\prime}}
}
\, ,  \nonumber
\end{eqnarray}
\begin{eqnarray} \label{R4_app}
{\cal T}^{[10], n+1}_{x_{~}^{~} x_{~}^{\prime} y_{~}^{~} y_{~}^{\prime}} = && \\
=\sum_{\substack{x_1^{~} x_2^{~} x_3^{~} x_4^{~} x_5^{~} x_6^{~} \\ x_1^{\prime} x_2^{\prime} x_3^{\prime} x_4^{\prime} x_5^{\prime} x_6^{\prime} \\ y_1^{~} y_2^{~} y_3^{~} y_4^{~} y_5^{~} y_6^{~} \\ y_1^{\prime} y_2^{\prime} y_3^{\prime} y_4^{\prime} y_5^{\prime} y_6^{\prime}}}^{~}&&
\substack{
{\cal T}_{
\left(x_1^{~}x_2^{~}x_3^{~}x_4^{~}\right) 
\left(x_{1}^{\prime}x_{2}^{\prime}x_{3}^{\prime}x_{4}^{\prime}\right) 
\left(y_1^{~}y_2^{~}y_3^{~}y_4^{~}\right) 
\left(y_{1}^{\prime}y_{2}^{\prime}y_{3}^{\prime}y_{4}^{\prime}\right)}^{[10], {n+1}} \\
U^{2}_{1, (x_1^{~} x_2^{~}) x_5^{~}} 
U^{2}_{2, (x_3^{~} x_4^{~}) x_6^{~}} 
U^{2}_{5, (x_5^{~} x_6^{~}) x_{~}^{~}} \\
U^{2}_{1, (x_1^{\prime} x_2^{\prime}) x_5^{\prime}} 
U^{2}_{2, (x_3^{\prime} x_4^{\prime}) x_6^{\prime}} 
U^{2}_{5, (x_5^{\prime} x_6^{\prime}) x_{~}^{\prime}} \\
U^{2}_{3, (y_1^{~} y_2^{~}) y_5^{~}} 
U^{2}_{4, (y_3^{~} y_4^{~}) y_6^{~}} 
U^{2}_{6, (y_5^{~} y_6^{~}) y_{~}^{~}} \\
U^{2}_{3, (y_1^{\prime} y_2^{\prime}) y_5^{\prime}} 
U^{2}_{4, (y_3^{\prime} y_4^{\prime}) y_6^{\prime}} 
U^{2}_{6, (y_5^{\prime} y_6^{\prime}) y_{~}^{\prime}}
}
\, .  \nonumber
\end{eqnarray}
Next, we abbreviate the notation by omitting all the tensor indices except for the tensor-type index (in square brackets) and the iteration step $n$ (the omitted tensor indices are the same as in Eq.~\eqref{R1_app} in each of the formulas). 
For brevity, we also omit the repeated $U$ and list the corresponding indices in the form of the $4 \times 3$ matrices instead.

\begin{eqnarray}
{\cal T}^{[1], n+1}_{~} &\leftarrow&
\sum
{\cal T}_{~}^{[1], {n+1}}
U^{\left(
\substack{
{1,}^{~}_{{~}_{~}^{~} {~}_{~}^{~}}
{1,}^{~}_{{~}_{~}^{~} {~}_{~}^{~}}
{1}^{~}_{{~}_{~}^{~}} \\
{1,}^{~}_{{~}_{~}^{~} {~}_{~}^{~}}
{1,}^{~}_{{~}_{~}^{~} {~}_{~}^{~}}
{1}^{~}_{{~}_{~}^{~}} \\
{1,}^{~}_{{~}_{~}^{~} {~}_{~}^{~}}
{1,}^{~}_{{~}_{~}^{~} {~}_{~}^{~}}
{1}^{~}_{{~}_{~}^{~}} \\
{1,}^{~}_{{~}_{~}^{~} {~}_{~}^{~}}
{1,}^{~}_{{~}_{~}^{~} {~}_{~}^{~}}
{1}^{~}_{{~}_{~}^{~}}
}
\right)}_{~} \, ,  \nonumber
\\
{\cal T}^{[2], n+1}_{~} &\leftarrow&
\sum
{\cal T}_{~}^{[2], {n+1}}
U^{\left(
\substack{
{1,}^{~}_{{~}_{~}^{~} {~}_{~}^{~}}
{1,}^{~}_{{~}_{~}^{~} {~}_{~}^{~}}
{1}^{~}_{{~}_{~}^{~}} \\
{1,}^{~}_{{~}_{~}^{~} {~}_{~}^{~}}
{1,}^{~}_{{~}_{~}^{~} {~}_{~}^{~}}
{1}^{~}_{{~}_{~}^{~}} \\
{2,}^{~}_{{~}_{~}^{~} {~}_{~}^{~}}
{2,}^{~}_{{~}_{~}^{~} {~}_{~}^{~}}
{2}^{~}_{{~}_{~}^{~}} \\
{1,}^{~}_{{~}_{~}^{~} {~}_{~}^{~}}
{1,}^{~}_{{~}_{~}^{~} {~}_{~}^{~}}
{1}^{~}_{{~}_{~}^{~}}
}
\right)}_{~} \, ,  \nonumber
\\
{\cal T}^{[3], n+1}_{~} &\leftarrow&
\sum
{\cal T}_{~}^{[3], {n+1}}
U^{\left(
\substack{
{1,}^{~}_{{~}_{~}^{~} {~}_{~}^{~}}
{1,}^{~}_{{~}_{~}^{~} {~}_{~}^{~}}
{1}^{~}_{{~}_{~}^{~}} \\
{2,}^{~}_{{~}_{~}^{~} {~}_{~}^{~}}
{2,}^{~}_{{~}_{~}^{~} {~}_{~}^{~}}
{2}^{~}_{{~}_{~}^{~}} \\
{1,}^{~}_{{~}_{~}^{~} {~}_{~}^{~}}
{1,}^{~}_{{~}_{~}^{~} {~}_{~}^{~}}
{1}^{~}_{{~}_{~}^{~}} \\
{1,}^{~}_{{~}_{~}^{~} {~}_{~}^{~}}
{1,}^{~}_{{~}_{~}^{~} {~}_{~}^{~}}
{1}^{~}_{{~}_{~}^{~}}
}
\right)}_{~} \, ,  \nonumber
\\
{\cal T}^{[4], n+1}_{~} &\leftarrow&
\sum
{\cal T}_{~}^{[4], {n+1}}
U^{\left(
\substack{
{1,}^{~}_{{~}_{~}^{~} {~}_{~}^{~}}
{1,}^{~}_{{~}_{~}^{~} {~}_{~}^{~}}
{1}^{~}_{{~}_{~}^{~}} \\
{1,}^{~}_{{~}_{~}^{~} {~}_{~}^{~}}
{1,}^{~}_{{~}_{~}^{~} {~}_{~}^{~}}
{1}^{~}_{{~}_{~}^{~}} \\
{1,}^{~}_{{~}_{~}^{~} {~}_{~}^{~}}
{1,}^{~}_{{~}_{~}^{~} {~}_{~}^{~}}
{1}^{~}_{{~}_{~}^{~}} \\
{2,}^{~}_{{~}_{~}^{~} {~}_{~}^{~}}
{2,}^{~}_{{~}_{~}^{~} {~}_{~}^{~}}
{2}^{~}_{{~}_{~}^{~}}
}
\right)}_{~} \, ,  \nonumber
\\
{\cal T}^{[5], n+1}_{~} &\leftarrow&
\sum
{\cal T}_{~}^{[5], {n+1}}
U^{\left(
\substack{
{2,}^{~}_{{~}_{~}^{~} {~}_{~}^{~}}
{2,}^{~}_{{~}_{~}^{~} {~}_{~}^{~}}
{2}^{~}_{{~}_{~}^{~}} \\
{1,}^{~}_{{~}_{~}^{~} {~}_{~}^{~}}
{1,}^{~}_{{~}_{~}^{~} {~}_{~}^{~}}
{1}^{~}_{{~}_{~}^{~}} \\
{1,}^{~}_{{~}_{~}^{~} {~}_{~}^{~}}
{1,}^{~}_{{~}_{~}^{~} {~}_{~}^{~}}
{1}^{~}_{{~}_{~}^{~}} \\
{1,}^{~}_{{~}_{~}^{~} {~}_{~}^{~}}
{1,}^{~}_{{~}_{~}^{~} {~}_{~}^{~}}
{1}^{~}_{{~}_{~}^{~}}
}
\right)}_{~} \, ,  \nonumber
\\
{\cal T}^{[6], n+1}_{~} &\leftarrow&
\sum
{\cal T}_{~}^{[6], {n+1}}
U^{\left(
\substack{
{1,}^{~}_{{~}_{~}^{~} {~}_{~}^{~}}
{1,}^{~}_{{~}_{~}^{~} {~}_{~}^{~}}
{1}^{~}_{{~}_{~}^{~}} \\
{2,}^{~}_{{~}_{~}^{~} {~}_{~}^{~}}
{2,}^{~}_{{~}_{~}^{~} {~}_{~}^{~}}
{2}^{~}_{{~}_{~}^{~}} \\
{2,}^{~}_{{~}_{~}^{~} {~}_{~}^{~}}
{2,}^{~}_{{~}_{~}^{~} {~}_{~}^{~}}
{2}^{~}_{{~}_{~}^{~}} \\
{1,}^{~}_{{~}_{~}^{~} {~}_{~}^{~}}
{1,}^{~}_{{~}_{~}^{~} {~}_{~}^{~}}
{1}^{~}_{{~}_{~}^{~}}
}
\right)}_{~} \, ,  \nonumber
\\
{\cal T}^{[7], n+1}_{~} &\leftarrow&
\sum
{\cal T}_{~}^{[7], {n+1}}
U^{\left(
\substack{
{1,}^{~}_{{~}_{~}^{~} {~}_{~}^{~}}
{1,}^{~}_{{~}_{~}^{~} {~}_{~}^{~}}
{1}^{~}_{{~}_{~}^{~}} \\
{2,}^{~}_{{~}_{~}^{~} {~}_{~}^{~}}
{2,}^{~}_{{~}_{~}^{~} {~}_{~}^{~}}
{2}^{~}_{{~}_{~}^{~}} \\
{1,}^{~}_{{~}_{~}^{~} {~}_{~}^{~}}
{1,}^{~}_{{~}_{~}^{~} {~}_{~}^{~}}
{1}^{~}_{{~}_{~}^{~}} \\
{2,}^{~}_{{~}_{~}^{~} {~}_{~}^{~}}
{2,}^{~}_{{~}_{~}^{~} {~}_{~}^{~}}
{2}^{~}_{{~}_{~}^{~}}
}
\right)}_{~} \, ,  \nonumber
\\
{\cal T}^{[8], n+1}_{~} &\leftarrow&
\sum
{\cal T}_{~}^{[8], {n+1}}
U^{\left(
\substack{
{2,}^{~}_{{~}_{~}^{~} {~}_{~}^{~}}
{2,}^{~}_{{~}_{~}^{~} {~}_{~}^{~}}
{2}^{~}_{{~}_{~}^{~}} \\
{1,}^{~}_{{~}_{~}^{~} {~}_{~}^{~}}
{1,}^{~}_{{~}_{~}^{~} {~}_{~}^{~}}
{1}^{~}_{{~}_{~}^{~}} \\
{1,}^{~}_{{~}_{~}^{~} {~}_{~}^{~}}
{1,}^{~}_{{~}_{~}^{~} {~}_{~}^{~}}
{1}^{~}_{{~}_{~}^{~}} \\
{2,}^{~}_{{~}_{~}^{~} {~}_{~}^{~}}
{2,}^{~}_{{~}_{~}^{~} {~}_{~}^{~}}
{2}^{~}_{{~}_{~}^{~}}
}
\right)}_{~} \, ,  \nonumber
\\
{\cal T}^{[9], n+1}_{~} &\leftarrow&
\sum
{\cal T}_{~}^{[9], {n+1}}
U^{\left(
\substack{
{2,}^{~}_{{~}_{~}^{~} {~}_{~}^{~}}
{2,}^{~}_{{~}_{~}^{~} {~}_{~}^{~}}
{2}^{~}_{{~}_{~}^{~}} \\
{1,}^{~}_{{~}_{~}^{~} {~}_{~}^{~}}
{1,}^{~}_{{~}_{~}^{~} {~}_{~}^{~}}
{1}^{~}_{{~}_{~}^{~}} \\
{2,}^{~}_{{~}_{~}^{~} {~}_{~}^{~}}
{2,}^{~}_{{~}_{~}^{~} {~}_{~}^{~}}
{2}^{~}_{{~}_{~}^{~}} \\
{1,}^{~}_{{~}_{~}^{~} {~}_{~}^{~}}
{1,}^{~}_{{~}_{~}^{~} {~}_{~}^{~}}
{1}^{~}_{{~}_{~}^{~}}
}
\right)}_{~} \, ,  \nonumber
\\
{\cal T}^{[10], n+1}_{~} &\leftarrow&
\sum
{\cal T}_{~}^{[10], {n+1}}
U^{\left(
\substack{
{2,}^{~}_{{~}_{~}^{~} {~}_{~}^{~}}
{2,}^{~}_{{~}_{~}^{~} {~}_{~}^{~}}
{2}^{~}_{{~}_{~}^{~}} \\
{2,}^{~}_{{~}_{~}^{~} {~}_{~}^{~}}
{2,}^{~}_{{~}_{~}^{~} {~}_{~}^{~}}
{2}^{~}_{{~}_{~}^{~}} \\
{2,}^{~}_{{~}_{~}^{~} {~}_{~}^{~}}
{2,}^{~}_{{~}_{~}^{~} {~}_{~}^{~}}
{2}^{~}_{{~}_{~}^{~}} \\
{2,}^{~}_{{~}_{~}^{~} {~}_{~}^{~}}
{2,}^{~}_{{~}_{~}^{~} {~}_{~}^{~}}
{2}^{~}_{{~}_{~}^{~}}
}
\right)}_{~} \, .  \nonumber
\end{eqnarray}

\section{Extensions of impurity tensors}

\subsection{Magnetization} \label{Magnetization_app}

The extension of the impurity tensor $\tilde{\cal T}^{n}$ is performed by taking an average over four central spins of the impurity in the extension pattern ${\cal T}^{[1]}$  (cf.~Eq.~(\ref{Eq_Crazy_app}))
\begin{eqnarray} \label{extension_appX}
\tilde{\cal T}_{
\left(x_1^{~}x_2^{~}x_3^{~}x_4^{~}\right) 
\left(x_{1}^{\prime}x_{2}^{\prime}x_{3}^{\prime}x_{4}^{\prime}\right) 
\left(y_1^{~}y_2^{~}y_3^{~}y_4^{~}\right) 
\left(y_{1}^{\prime}y_{2}^{\prime}y_{3}^{\prime}y_{4}^{\prime}\right)}^{n+1} = \, \, \nonumber \\
\dfrac{1}{4} \sum_{\substack{a b c d e f g h i \\ j k l m n^{\prime} o p q \\ r s t u v w x}}^{~}
\left[
\left(
{{{{\cal T}^{[10], n}_{x_1^{~} a^{~}_{~} y_1^{~} d^{~}_{~}}  
{\cal T}^{[5], n}_{a^{~}_{~} b^{~}_{~} y_{2}^{~} e^{~}_{~}}
{\cal T}^{[3], n}_{b^{~}_{~} c^{~}_{~} y_{3}^{~} f^{~}_{~}} 
{\cal T}^{[10], n}_{c^{~}_{~} x_{1}^{\prime} y_{4}^{~} g^{~}_{~}} } \atop
{{\cal T}^{[2], n}_{x_{2}^{~} h^{~}_{~} d^{~}_{~} k^{~}_{~}}
 \mathbf{\tilde{\cal T}}^{n}_{h^{~}_{~} i^{~}_{~} e^{~}_{~} l^{~}_{~}}
{\cal T}^{[1], n}_{i^{~}_{~} j^{~}_{~} f^{~}_{~} m^{~}_{~}} 
{\cal T}^{[2], n}_{j^{~}_{~} x_{2}^{\prime} g_{~}^{~} n^{\prime}_{~}}}} \atop
{{{\cal T}^{[4], n}_{x_{3}^{~} o_{~}^{~} k_{~}^{~} r_{~}^{~}}  
  {\cal T}^{[1], n}_{o_{~}^{~} p_{~}^{~} l_{~}^{~} s_{~}^{~}} 
  {\cal T}^{[1], n}_{p_{~}^{~} q_{~}^{~} m_{~}^{~} t_{~}^{~}} 
  {\cal T}^{[4], n}_{q_{~}^{~} x_3^{\prime} n_{~}^{\prime} u_{~}^{~}} } \atop
{{\cal T}^{[10], n}_{x_4^{~} v_{~}^{~} r_{~}^{~} y_1^{\prime}}
 {\cal T}^{[5], n}_{v_{~}^{~} w_{~}^{~} s_{~}^{~} y_{2}^{\prime}}
 {\cal T}^{[3], n}_{w_{~}^{~} x_{~}^{~} t_{~}^{~} y_3^{\prime}}
 {\cal T}^{[10], n}_{x_{~}^{~} x_4^{\prime} u_{~}^{~} y_4^{\prime}}}}} 
 \right) 
 \right. 
 + \nonumber
 \\
\left(
{{{{\cal T}^{[10], n}_{x_1^{~} a^{~}_{~} y_1^{~} d^{~}_{~}}  
{\cal T}^{[5], n}_{a^{~}_{~} b^{~}_{~} y_{2}^{~} e^{~}_{~}}
{\cal T}^{[3], n}_{b^{~}_{~} c^{~}_{~} y_{3}^{~} f^{~}_{~}} 
{\cal T}^{[10], n}_{c^{~}_{~} x_{1}^{\prime} y_{4}^{~} g^{~}_{~}} } \atop
{{\cal T}^{[2], n}_{x_{2}^{~} h^{~}_{~} d^{~}_{~} k^{~}_{~}}
 {\cal T}^{[1], n}_{h^{~}_{~} i^{~}_{~} e^{~}_{~} l^{~}_{~}}
\mathbf{\tilde{\cal T}}^{n}_{i^{~}_{~} j^{~}_{~} f^{~}_{~} m^{~}_{~}} 
{\cal T}^{[2], n}_{j^{~}_{~} x_{2}^{\prime} g_{~}^{~} n^{\prime}_{~}}}} \atop
{{{\cal T}^{[4], n}_{x_{3}^{~} o_{~}^{~} k_{~}^{~} r_{~}^{~}}  
  {\cal T}^{[1], n}_{o_{~}^{~} p_{~}^{~} l_{~}^{~} s_{~}^{~}} 
  {\cal T}^{[1], n}_{p_{~}^{~} q_{~}^{~} m_{~}^{~} t_{~}^{~}} 
  {\cal T}^{[4], n}_{q_{~}^{~} x_3^{\prime} n_{~}^{\prime} u_{~}^{~}} } \atop
{{\cal T}^{[10], n}_{x_4^{~} v_{~}^{~} r_{~}^{~} y_1^{\prime}}
 {\cal T}^{[5], n}_{v_{~}^{~} w_{~}^{~} s_{~}^{~} y_{2}^{\prime}}
 {\cal T}^{[3], n}_{w_{~}^{~} x_{~}^{~} t_{~}^{~} y_3^{\prime}}
 {\cal T}^{[10], n}_{x_{~}^{~} x_4^{\prime} u_{~}^{~} y_4^{\prime}}}}} 
 \right) 
 +  \nonumber
  \\
  \left(
{{{{\cal T}^{[10], n}_{x_1^{~} a^{~}_{~} y_1^{~} d^{~}_{~}}  
{\cal T}^{[5], n}_{a^{~}_{~} b^{~}_{~} y_{2}^{~} e^{~}_{~}}
{\cal T}^{[3], n}_{b^{~}_{~} c^{~}_{~} y_{3}^{~} f^{~}_{~}} 
{\cal T}^{[10], n}_{c^{~}_{~} x_{1}^{\prime} y_{4}^{~} g^{~}_{~}} } \atop
{{\cal T}^{[2], n}_{x_{2}^{~} h^{~}_{~} d^{~}_{~} k^{~}_{~}}
 {\cal T}^{[1], n}_{h^{~}_{~} i^{~}_{~} e^{~}_{~} l^{~}_{~}}
{\cal T}^{[1], n}_{i^{~}_{~} j^{~}_{~} f^{~}_{~} m^{~}_{~}} 
{\cal T}^{[2], n}_{j^{~}_{~} x_{2}^{\prime} g_{~}^{~} n^{\prime}_{~}}}} \atop
{{{\cal T}^{[4], n}_{x_{3}^{~} o_{~}^{~} k_{~}^{~} r_{~}^{~}}  
  {\cal T}^{[1], n}_{o_{~}^{~} p_{~}^{~} l_{~}^{~} s_{~}^{~}} 
  \mathbf{\tilde{\cal T}}^{n}_{p_{~}^{~} q_{~}^{~} m_{~}^{~} t_{~}^{~}} 
  {\cal T}^{[4], n}_{q_{~}^{~} x_3^{\prime} n_{~}^{\prime} u_{~}^{~}} } \atop
{{\cal T}^{[10], n}_{x_4^{~} v_{~}^{~} r_{~}^{~} y_1^{\prime}}
 {\cal T}^{[5], n}_{v_{~}^{~} w_{~}^{~} s_{~}^{~} y_{2}^{\prime}}
 {\cal T}^{[3], n}_{w_{~}^{~} x_{~}^{~} t_{~}^{~} y_3^{\prime}}
 {\cal T}^{[10], n}_{x_{~}^{~} x_4^{\prime} u_{~}^{~} y_4^{\prime}}}}} 
\right) 
 +  \nonumber
  \\
\left. 
\left(
{{{{\cal T}^{[10], n}_{x_1^{~} a^{~}_{~} y_1^{~} d^{~}_{~}}  
{\cal T}^{[5], n}_{a^{~}_{~} b^{~}_{~} y_{2}^{~} e^{~}_{~}}
{\cal T}^{[3], n}_{b^{~}_{~} c^{~}_{~} y_{3}^{~} f^{~}_{~}} 
{\cal T}^{[10], n}_{c^{~}_{~} x_{1}^{\prime} y_{4}^{~} g^{~}_{~}} } \atop
{{\cal T}^{[2], n}_{x_{2}^{~} h^{~}_{~} d^{~}_{~} k^{~}_{~}}
 {\cal T}^{[1], n}_{h^{~}_{~} i^{~}_{~} e^{~}_{~} l^{~}_{~}}
{\cal T}^{[1], n}_{i^{~}_{~} j^{~}_{~} f^{~}_{~} m^{~}_{~}} 
{\cal T}^{[2], n}_{j^{~}_{~} x_{2}^{\prime} g_{~}^{~} n^{\prime}_{~}}}} \atop
{{{\cal T}^{[4], n}_{x_{3}^{~} o_{~}^{~} k_{~}^{~} r_{~}^{~}}  
  \mathbf{\tilde{\cal T}}^{n}_{o_{~}^{~} p_{~}^{~} l_{~}^{~} s_{~}^{~}} 
  {\cal T}^{[1], n}_{p_{~}^{~} q_{~}^{~} m_{~}^{~} t_{~}^{~}} 
  {\cal T}^{[4], n}_{q_{~}^{~} x_3^{\prime} n_{~}^{\prime} u_{~}^{~}} } \atop
{{\cal T}^{[10], n}_{x_4^{~} v_{~}^{~} r_{~}^{~} y_1^{\prime}}
 {\cal T}^{[5], n}_{v_{~}^{~} w_{~}^{~} s_{~}^{~} y_{2}^{\prime}}
 {\cal T}^{[3], n}_{w_{~}^{~} x_{~}^{~} t_{~}^{~} y_3^{\prime}}
 {\cal T}^{[10], n}_{x_{~}^{~} x_4^{\prime} u_{~}^{~} y_4^{\prime}}}}} 
 \right)
 \right] 
 \, ,  \nonumber
\end{eqnarray}
where the letter $\mathbf{\tilde{\cal T}}$ in bold is meant to stress the readability of the above formula.

\subsection{Bond energy} \label{Bond_energy_app}
The expression for the bond energy begins with the identical initial impurity tensor, as we used in the spontaneous magnetization (see Eq.~{\eqref{mag_imp_init}}). However, the averaging over the bond energy, corresponding to the spin-spin pairs, is necessary to be performed. Therefore, the first extension takes an average over the four different neighboring pairs of the impurities in the extension pattern ${\cal T}^{[1]}$
\begin{eqnarray}
{\tilde{\tilde{\cal T}}}_{
\left(x_1^{~}x_2^{~}x_3^{~}x_4^{~}\right) 
\left(x_{1}^{\prime}x_{2}^{\prime}x_{3}^{\prime}x_{4}^{\prime}\right) 
\left(y_1^{~}y_2^{~}y_3^{~}y_4^{~}\right) 
\left(y_{1}^{\prime}y_{2}^{\prime}y_{3}^{\prime}y_{4}^{\prime}\right)}^{n=1} = \, \, \nonumber \\
\dfrac{1}{4} \sum_{\substack{a b c d e f g h i \\ j k l m n^{\prime} o p q \\ r s t u v w x}}^{~}
\left[
\left(
{{{{\cal T}^{[10], n=0}_{x_1^{~} a^{~}_{~} y_1^{~} d^{~}_{~}}  
{\cal T}^{[5], n=0}_{a^{~}_{~} b^{~}_{~} y_{2}^{~} e^{~}_{~}}
{\cal T}^{[3], n=0}_{b^{~}_{~} c^{~}_{~} y_{3}^{~} f^{~}_{~}} 
{\cal T}^{[10], n=0}_{c^{~}_{~} x_{1}^{\prime} y_{4}^{~} g^{~}_{~}} } \atop
{{\cal T}^{[2], n=0}_{x_{2}^{~} h^{~}_{~} d^{~}_{~} k^{~}_{~}}
 \mathbf{\tilde{\cal T}}^{n=0}_{h^{~}_{~} i^{~}_{~} e^{~}_{~} l^{~}_{~}}
 \mathbf{\tilde{\cal T}}^{n=0}_{i^{~}_{~} j^{~}_{~} f^{~}_{~} m^{~}_{~}} 
{\cal T}^{[2], n=0}_{j^{~}_{~} x_{2}^{\prime} g_{~}^{~} n^{\prime}_{~}}}} \atop
{{{\cal T}^{[4], n=0}_{x_{3}^{~} o_{~}^{~} k_{~}^{~} r_{~}^{~}}  
  {\cal T}^{[1], n=0}_{o_{~}^{~} p_{~}^{~} l_{~}^{~} s_{~}^{~}} 
  {\cal T}^{[1], n=0}_{p_{~}^{~} q_{~}^{~} m_{~}^{~} t_{~}^{~}} 
  {\cal T}^{[4], n=0}_{q_{~}^{~} x_3^{\prime} n_{~}^{\prime} u_{~}^{~}} } \atop
{{\cal T}^{[10], n=0}_{x_4^{~} v_{~}^{~} r_{~}^{~} y_1^{\prime}}
 {\cal T}^{[5], n=0}_{v_{~}^{~} w_{~}^{~} s_{~}^{~} y_{2}^{\prime}}
 {\cal T}^{[3], n=0}_{w_{~}^{~} x_{~}^{~} t_{~}^{~} y_3^{\prime}}
 {\cal T}^{[10], n=0}_{x_{~}^{~} x_4^{\prime} u_{~}^{~} y_4^{\prime}}}}} 
 \right) 
 \right. + \nonumber
 \\
\left(
{{{{\cal T}^{[10], n=0}_{x_1^{~} a^{~}_{~} y_1^{~} d^{~}_{~}}  
{\cal T}^{[5], n=0}_{a^{~}_{~} b^{~}_{~} y_{2}^{~} e^{~}_{~}}
{\cal T}^{[3], n=0}_{b^{~}_{~} c^{~}_{~} y_{3}^{~} f^{~}_{~}} 
{\cal T}^{[10], n=0}_{c^{~}_{~} x_{1}^{\prime} y_{4}^{~} g^{~}_{~}} } \atop
{{\cal T}^{[2], n=0}_{x_{2}^{~} h^{~}_{~} d^{~}_{~} k^{~}_{~}}
 {\cal T}^{[1], n=0}_{h^{~}_{~} i^{~}_{~} e^{~}_{~} l^{~}_{~}}
\mathbf{\tilde{\cal T}}^{n=0}_{i^{~}_{~} j^{~}_{~} f^{~}_{~} m^{~}_{~}} 
{\cal T}^{[2], n=0}_{j^{~}_{~} x_{2}^{\prime} g_{~}^{~} n^{\prime}_{~}}}} \atop
{{{\cal T}^{[4], n=0}_{x_{3}^{~} o_{~}^{~} k_{~}^{~} r_{~}^{~}}  
  {\cal T}^{[1], n=0}_{o_{~}^{~} p_{~}^{~} l_{~}^{~} s_{~}^{~}} 
  \mathbf{\tilde{\cal T}}^{n=0}_{p_{~}^{~} q_{~}^{~} m_{~}^{~} t_{~}^{~}} 
  {\cal T}^{[4], n=0}_{q_{~}^{~} x_3^{\prime} n_{~}^{\prime} u_{~}^{~}} } \atop
{{\cal T}^{[10], n=0}_{x_4^{~} v_{~}^{~} r_{~}^{~} y_1^{\prime}}
 {\cal T}^{[5], n=0}_{v_{~}^{~} w_{~}^{~} s_{~}^{~} y_{2}^{\prime}}
 {\cal T}^{[3], n=0}_{w_{~}^{~} x_{~}^{~} t_{~}^{~} y_3^{\prime}}
 {\cal T}^{[10], n=0}_{x_{~}^{~} x_4^{\prime} u_{~}^{~} y_4^{\prime}}}}} 
\right) 
 +  \nonumber
  \\
\left(
{{{{\cal T}^{[10], n=0}_{x_1^{~} a^{~}_{~} y_1^{~} d^{~}_{~}}  
{\cal T}^{[5], n=0}_{a^{~}_{~} b^{~}_{~} y_{2}^{~} e^{~}_{~}}
{\cal T}^{[3], n=0}_{b^{~}_{~} c^{~}_{~} y_{3}^{~} f^{~}_{~}} 
{\cal T}^{[10], n=0}_{c^{~}_{~} x_{1}^{\prime} y_{4}^{~} g^{~}_{~}} } \atop
{{\cal T}^{[2], n=0}_{x_{2}^{~} h^{~}_{~} d^{~}_{~} k^{~}_{~}}
 {\cal T}^{[1], n=0}_{h^{~}_{~} i^{~}_{~} e^{~}_{~} l^{~}_{~}}
{\cal T}^{[1], n=0}_{i^{~}_{~} j^{~}_{~} f^{~}_{~} m^{~}_{~}} 
{\cal T}^{[2], n=0}_{j^{~}_{~} x_{2}^{\prime} g_{~}^{~} n^{\prime}_{~}}}} \atop
{{{\cal T}^{[4], n=0}_{x_{3}^{~} o_{~}^{~} k_{~}^{~} r_{~}^{~}}  
  \mathbf{\tilde{\cal T}}^{n=0}_{o_{~}^{~} p_{~}^{~} l_{~}^{~} s_{~}^{~}} 
  \mathbf{\tilde{\cal T}}^{n=0}_{p_{~}^{~} q_{~}^{~} m_{~}^{~} t_{~}^{~}} 
  {\cal T}^{[4], n=0}_{q_{~}^{~} x_3^{\prime} n_{~}^{\prime} u_{~}^{~}} } \atop
{{\cal T}^{[10], n=0}_{x_4^{~} v_{~}^{~} r_{~}^{~} y_1^{\prime}}
 {\cal T}^{[5], n=0}_{v_{~}^{~} w_{~}^{~} s_{~}^{~} y_{2}^{\prime}}
 {\cal T}^{[3], n=0}_{w_{~}^{~} x_{~}^{~} t_{~}^{~} y_3^{\prime}}
 {\cal T}^{[10], n=0}_{x_{~}^{~} x_4^{\prime} u_{~}^{~} y_4^{\prime}}}}} 
\right) 
 +  \nonumber
  \\
  \left. 
  \left(
{{{{\cal T}^{[10], n=0}_{x_1^{~} a^{~}_{~} y_1^{~} d^{~}_{~}}  
{\cal T}^{[5], n=0}_{a^{~}_{~} b^{~}_{~} y_{2}^{~} e^{~}_{~}}
{\cal T}^{[3], n=0}_{b^{~}_{~} c^{~}_{~} y_{3}^{~} f^{~}_{~}} 
{\cal T}^{[10], n=0}_{c^{~}_{~} x_{1}^{\prime} y_{4}^{~} g^{~}_{~}} } \atop
{{\cal T}^{[2], n=0}_{x_{2}^{~} h^{~}_{~} d^{~}_{~} k^{~}_{~}}
 \mathbf{\tilde{\cal T}}^{n=0}_{h^{~}_{~} i^{~}_{~} e^{~}_{~} l^{~}_{~}}
{\cal T}^{[1], n=0}_{i^{~}_{~} j^{~}_{~} f^{~}_{~} m^{~}_{~}} 
{\cal T}^{[2], n=0}_{j^{~}_{~} x_{2}^{\prime} g_{~}^{~} n^{\prime}_{~}}}} \atop
{{{\cal T}^{[4], n=0}_{x_{3}^{~} o_{~}^{~} k_{~}^{~} r_{~}^{~}}  
  \mathbf{\tilde{\cal T}}^{n=0}_{o_{~}^{~} p_{~}^{~} l_{~}^{~} s_{~}^{~}} 
  {\cal T}^{[1], n=0}_{p_{~}^{~} q_{~}^{~} m_{~}^{~} t_{~}^{~}} 
  {\cal T}^{[4], n=0}_{q_{~}^{~} x_3^{\prime} n_{~}^{\prime} u_{~}^{~}} } \atop
{{\cal T}^{[10], n=0}_{x_4^{~} v_{~}^{~} r_{~}^{~} y_1^{\prime}}
 {\cal T}^{[5], n=0}_{v_{~}^{~} w_{~}^{~} s_{~}^{~} y_{2}^{\prime}}
 {\cal T}^{[3], n=0}_{w_{~}^{~} x_{~}^{~} t_{~}^{~} y_3^{\prime}}
 {\cal T}^{[10], n=0}_{x_{~}^{~} x_4^{\prime} u_{~}^{~} y_4^{\prime}}}}} 
 \right)
 \right]
 \, ,  \nonumber
\end{eqnarray}
If $n>0$, we proceed in the bond-energy extension of the impurity tensor according to the relation defined for magnetization in Appendix~\ref{Magnetization_app}.

\end{document}